\documentclass[sigconf,nonacm,dvipsnames]{acmart}

\pdfoutput=1

\usepackage{comment}
\usepackage{soul}
\usepackage{enumitem}
\usepackage{array}
\usepackage{makecell}

\setcopyright{acmlicensed}
\copyrightyear{2018}
\acmYear{2018}
\acmDOI{XXXXXXX.XXXXXXX}

\acmConference[Conference acronym 'XX]{Make sure to enter the correct
  conference title from your rights confirmation emai}{June 03--05,
  2018}{Woodstock, NY}
\acmISBN{978-1-4503-XXXX-X/18/06}

\begin{document}



\title{\sys{}: Recursively Expandable Abstracts for Directed Information Retrieval over Scientific Papers}

\author{Raymond Fok}
\email{rayfok@cs.washington.edu}
\affiliation{
  \institution{University of Washington}
  \city{Seattle}
  \state{WA}
  \country{USA}
}
\authornote{Work completed during an internship at Semantic Scholar, Allen Institute for AI.}

\author{Joseph Chee Chang}
\email{josephc@allenai.org}
\affiliation{
  \institution{Allen Institute for AI}
  \city{Seattle}
  \state{WA}
  \country{USA}
}

\author{Tal August}
\email{tala@allenai.org}
\affiliation{
  \institution{Allen Institute for AI}
  \city{Seattle}
  \state{WA}
  \country{USA}
}

\author{Amy X. Zhang}
\email{axz@cs.uw.edu}
\affiliation{
  \institution{University of Washington}
  \city{Seattle}
  \state{WA}
  \country{USA}
}

\author{Daniel S. Weld}
\email{danw@allenai.org}
\affiliation{
  \institution{Allen Institute for AI \&\ \\ University of Washington}
  \city{Seattle}
  \state{WA}
  \country{USA}
}

\renewcommand{\shortauthors}{}

\newcommand\ray[1]{\textcolor{blue}{[Ray]: #1}}
\newcommand\dw[1]{\textcolor{red}{[Dan]: #1}}
\newcommand\amy[1]{\textcolor{brown}{[Amy]: #1}}
\newcommand\joseph[1]{\textcolor{WildStrawberry}{[Joseph]: #1}}
\newcommand\jc[1]{\textcolor{WildStrawberry}{[Joseph]: #1}}
\newcommand\tal[1]{\textcolor{teal}{[Tal]: #1}}

\newcommand{\bug}{\mbox{\rule{2mm}{2mm}}}
\newcommand{\DSW}[1]{\bug \footnote{\textcolor{red}{\textit{DSW: #1}}}}

\newcommand{\sys}{\textsc{Qlarify}}

\newcolumntype{C}[1]{>{\centering\arraybackslash}p{#1}}

\begin{abstract}
Navigating the vast scientific literature often starts with browsing a paper’s abstract. However, when a reader seeks additional information, not present in the abstract, they face a costly cognitive chasm during their dive into the full text. To bridge this gap, we introduce \emph{recursively expandable abstracts}, a novel interaction paradigm that dynamically expands abstracts by progressively incorporating additional information from the papers’ full text. This lightweight interaction allows scholars to specify their information needs by quickly brushing over the abstract or selecting AI-suggested expandable entities. Relevant information is synthesized using a retrieval-augmented generation approach, presented as a fluid, threaded expansion of the abstract, and made efficiently verifiable via attribution to relevant source-passages in the paper. Through a series of user studies, we demonstrate the utility of recursively expandable abstracts and identify future opportunities to support low-effort and just-in-time exploration of long-form information contexts through LLM-powered interactions.

\end{abstract}




\begin{teaserfigure}
    \centering
    \includegraphics[width=0.95\textwidth]{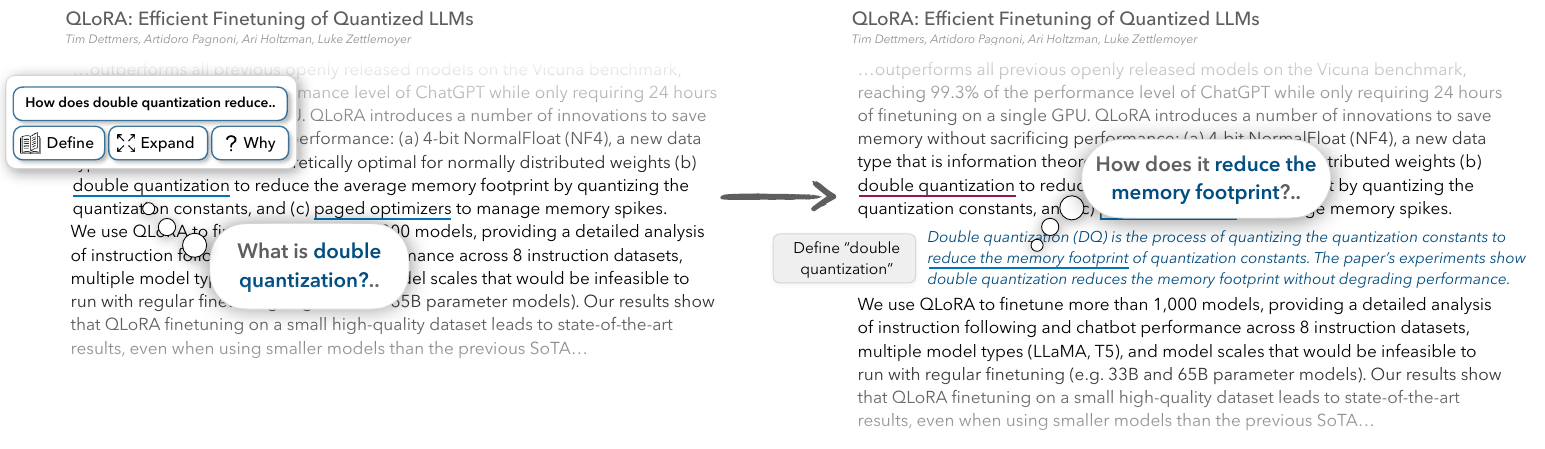}
    \caption{\textit{Recursively expandable abstracts} present a novel mixed-initiative interaction technique, leveraging large language models to enable a low-cost, on-demand, and fluid expansion of static abstracts with information retrieved from full papers.}
    \Description[]{}
    \label{fig:teaser}
\end{teaserfigure}

\maketitle

\section{Introduction}


\begin{teaserfigure}
    \centering
    \includegraphics[width=0.95\textwidth]{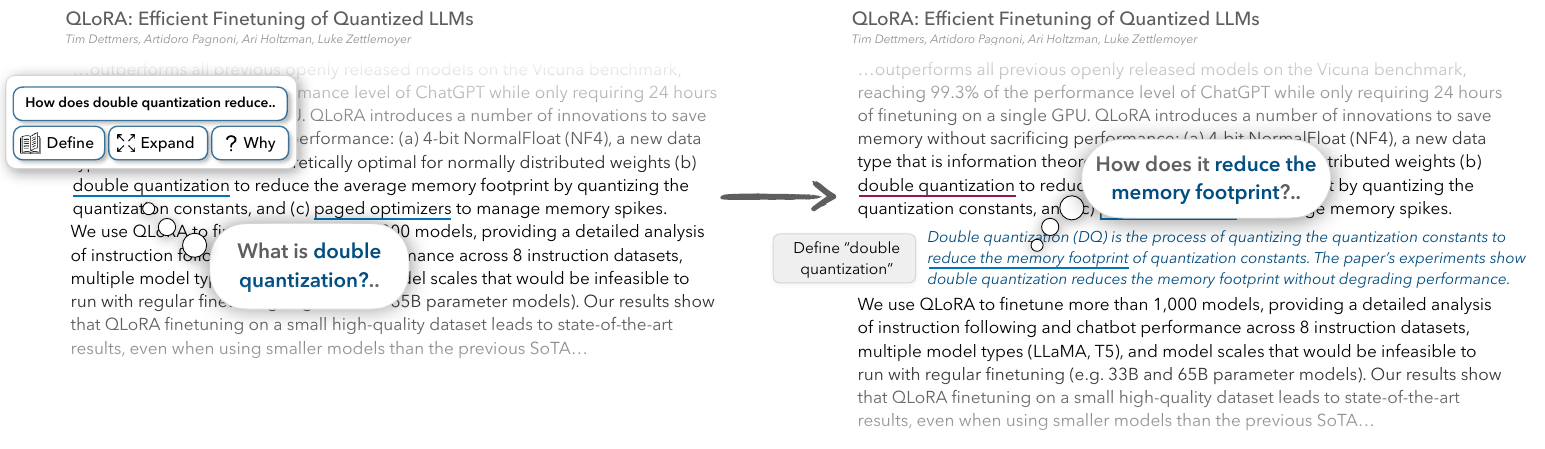}
    \caption{\textit{Recursively expandable abstracts} present a novel mixed-initiative interaction technique, leveraging large language models to enable a low-cost, on-demand, and fluid expansion of static abstracts with information retrieved from full papers.}
    \Description[]{}
    \label{fig:teaser}
\end{teaserfigure}

Reviewing prior literature is an important part of the scientific progress, helping scholars to survey what has already been done, identify open challenges, and spark inspiration~\cite{knopf_doing_2006}. 
In response, technological interventions have sought to aid scientists in discovering and consuming the vast literature. Systems such as Google Scholar
and Semantic Scholar
help researchers {\em discover} relevant papers, while other systems assist scholars' {\em consumption} with tools to aid comprehension~\cite{head_augmenting_2021, august_paperplain_2023}, efficiency~\cite{lee_spotlights_2016, fok_scim_2023}, or sensemaking~\cite{kang_synergi_2023, kang_threddy_2022} of these newly-found works.

Yet, at the seam of literature discovery and consumption there exists another challenge which has received less attention --- the preliminary, breadth-first exploration of papers. Scholars often need to browse {\em collections} of potentially related manuscripts, such as recurring email digests of recently published papers or proceedings of conferences. The exploratory triaging process allows scholars to assess papers for their relevance, identify future reading material, or satisfy an informational curiosity, without incurring the cognitive costs of reading each paper. A typical approach might start by reasoning over each paper's abstract to determine its relevance and legitimacy and examining only a subset in more detail~\cite{ishita_which_2018}. Literature discovery tools can provide scholars with potential papers and literature consumption tools can assist scholars in reading selected papers; however, neither set of tools assists in this preliminary exploration of papers through abstracts.

While often used for triage, abstracts have several limitations. As static summaries that condense upwards of 10,000 words into one or two paragraphs, abstracts rarely address all of a scholar's information needs and require clarification to fully comprehend. For instance, an abstract could mention the size of a study, but leave a scholar uncertain about the recruitment process, participant demographics, or analyses. This challenge of locating additional context is further complicated when scholars need to triage numerous papers. A scholar interested in understanding how prior work built on a specific dataset would need to manually open the PDFs for each candidate and search that paper for mentions of the dataset name in order to find passages that contained detailed descriptions about how the dataset was used.

Through a formative study, we identified common information needs that were often expressed as questions about the abstract and answerable with information from the paper's full text. We address these knowledge gaps by proposing a novel interaction technique, \textit{recursively expandable abstracts}. This technique bridges abstracts and full papers by enabling users to interactively ask clarifying questions and expand abstracts with information retrieved from the full paper texts. We reify expandable abstracts within \sys, an LLM-powered augmented reading interface for scientific paper abstracts. In contrast to other intelligent chat-based interactions with papers which require users to formulate questions from scratch, \sys{} showcases a dynamic text interaction paradigm where just-in-time questions are formed through direct, lightweight engagement with abstract itself.

By highlighting any part of an expandable abstract in \sys{}, users can request an elaboration of the abstract and progressively expand the abstract with relevant context from the full paper. These expandable abstracts are also recursive, affording clarifying questions that probe deeper into details of the paper in a threaded manner. \sys{} bootstraps the question-asking process by recommending potentially expandable entities within an abstract, providing information scent toward informative areas to expand (Figure~\ref{fig:teaser}). \sys{} reduces the cost of asking an appropriate question to a single click by providing two types of question scaffolding: (1) a contextually-sensitive AI-suggested question which aims to infer a user's information-seeking intent, and (2) three static questions---\textit{Define}, \textit{Expand}, and \textit{Why}---which mirror the common information needs scholars may desire in an expansion. Finally, \sys{} enables efficient verification of generated expansions with attribution via deep linking to relevant passages in the paper.

Though an interview study (N=9), field deployment (N=275), and comparative user study (N=12), we evaluated the design and efficacy of \sys{} and recursively expandable abstracts for abstract and paper exploration. In our studies, we found participants used the AI-suggested expandable entities as a guide for probing abstracts for details within papers, often creating threaded expansions to satisfy follow-up curiosities. Participants tended to ask more questions with \sys{} than a question answering baseline and overall preferred expandable abstracts over baseline paper exploration approaches. Our findings also indicate LLMs can effectively infer scholars' information-seeking intents within abstracts and add value by surfacing clarifying information generated from the full text of papers, with over 88\% accuracy using a standard retrieval-augmented generation approach over full papers. We conclude with future opportunities for recursive expansion interactions and their implications for AI-infused scholarly support tools.

In summary, we contribute the following:
\begin{itemize}
    \item A novel document-centered interaction technique, \textit{recursively expandable abstracts}, that allows users to progressively expand abstracts on-demand with information from a paper's full text.
    \item \sys, an augmented reading interface with LLM-enabled recursively expandable abstracts, aiding users in forming clarifying queries, extracting relevant information from papers, and validating the expanded content.
    \item Findings from an interview study (N=9), online deployment (N=275), and comparative user study (N=12) with scholars, characterizing the utility of recursively expandable abstracts for rapid and thorough interactive information foraging across papers.
\end{itemize}

\section{Related Work}

\subsection{Addressing Document-Centered Information Needs with Summarization}
The aim of summarization is to condense long documents into short and concise texts, encapsulating the most important information required for comprehension. With the advent of neural architectures~\cite{koh_empirical_2022}, significant efforts have been dedicated to improving the capabilities of automatic text summarization systems. Some work has focused especially on summarizing domain-specific and long-form documents, such as scientific text~\cite{cachola_tldr_2020, yasunaga_scisummnet_2019, ibrahim_altmami_automatic_2022}. Consuming a traditional summary requires little to no user effort, but the static medium of a summary makes it impossible to capture the personalized and nuanced information needs of every individual. Some research has explored ways to incorporate humans in the loop to generate more personalized summaries~\cite{zhang_concepteva_2023, shapira_interactive_2022, ghodratnama_adaptive_2021}, but once generated, these summaries lack the ability for iterative refinement, for instance to reflect updated information needs.

Prior work has explored hierarchical approaches to summarization that enable a reader to interactively specify the degree of depth they wish to explore. These approaches require the authoring of summaries at differing levels of depth (e.g., for summarizing large-scale online discussions~\cite{zhang_wikum_2017}, books~\cite{wu_recursively_2021}, or web documents~\cite{bernstein_hypertext_2009, nelson_literary_1983}), forming a static summary tree artifact that enables structured navigation between the summaries and long documents. 
These summaries are typically constructed in a bottom-up fashion, working recursively starting from the full text, so as to break down the work into more manageable chunks.
In contrast, we take a top-down approach; rather than condensing information to \textit{generate} summaries, we instead leverage automated techniques to \textit{expand} summaries---incrementally and recursively---to reveal relevant information from a long document on-demand. By constructing expandable summaries in the same direction that readers explore (i.e., by drilling down), the summary trees generated by our system are personalized to each reader's exploration, as opposed to having to conform to a single rigid structure.

This concept of expanding text to incorporate more context and resolve ambiguities is also related to decontextualization, an NLP task studying automated approaches for rewriting extractive excerpts divorced from a longer document to be interpretable out of context while preserving meaning~\cite{choi_decontextualization_2021, newman_controllable_2023}. The expansion interaction we explore can be seen as a form of interactive decontextualization, sensitive to users' personal information needs.

\subsection{Querying Documents with QA Systems}
People often express their information needs within documents as natural language questions~\cite{jahanbakhsh_understanding_2022, ter_hoeve_conversations_2020, ko_inquisitive_2020}, and ask follow-up questions when an initial answer is not sufficient~\cite{meng_followupqg_2023}. Question answering is a long-standing problem in NLP, including considerable efforts for answering questions over scientific documents~\cite{dasigi_dataset_2021, rogers_qa_2023, saad-falcon_pdftriage_2023}. Recent advances in LLMs have seen the proliferation of prototypes for conversational question answering over long documents (e.g., \textit{ChatDoc}~\cite{chatDoc}, \textit{ChatPDF}~\cite{chatPDF}). The expansion interaction we propose in this work is related to these systems, but differs in two main ways: first, we aim to integrate answers in-situ by expanding the summary at the point where a question was asked to maintain the flow of reading, and second, we introduce mixed-initiative interactions that reduce the cost of forming and asking questions. To enable these question-driven expansions, we implement a retriever-reader architecture with a dense representation-based retriever and a generative, LLM-powered reader.

Attributed question answering~\cite{bohnet_attributed_2023}, where AI models return evidence in addition to their answer, has gained recent interest due to the potential for hallucinations in LLM-generated answers~\cite{maynez_faithfulness_2020, ji_survey_2023}. Some work has sought to more closely integrate attribution with generation (e.g., with post-editing~\cite{gao_rarr_2023}), while systems such as \textit{GopherCite}~\cite{menick_teaching_2022}, \textit{WebGPT}~\cite{nakano_webgpt_2022}, and \textit{LaMDA}~\cite{thoppilan_lamda_2022} place the burden of fact verification on the user by providing URLs and excerpts as supporting evidence. We take a similar user-centric approach to attribution, providing simple affordances for users to drill down into excerpts and then the full paper from an LLM-generated response.

Prior conversational interfaces with documents often make two assumptions: that users know what to ask, and that the most natural interaction is for users to manually type their intended questions. Yet this is not always the case~\cite{august_paperplain_2023}. Instead, we propose that carefully crafted interactions and language understanding techniques can effectively infer the intent of users, reducing the cost of asking questions to a single click. This work explores the potential for intelligent support to not only answer users' questions, but also suggest questions prompting relevant entities to expand. Recent work has begun to explore the effectiveness of LLMs in this task of question generation, such as for generating follow-up and clarification questions~\cite{meng_followupqg_2023, ko_inquisitive_2020, kumar_clarq_2020}. As prompting LLMs can be challenging~\cite{zamfirescu-pereira_why_2023, mishra_promptaid_2023}, \sys{} obviates the need for users to craft their own prompts by providing a selection of common questions. These questions are grounded in a taxonomy of document-centered information-seeking questions people ask while reading a document~\cite{ko_inquisitive_2020}, and refined for scientific documents through our formative study.

\subsection{Supporting Scientific Literature Review}
Scholars need to triage more papers in less time now more than ever~\cite{landhuis_scientific_2016}, facing constant information overload from the growing number of publications~\cite{national_science_foundation_publications_2021, chu_slowed_2021}, shift from paper to digital publishing~\cite{liu_reading_2005, tenopir_electronic_2009}, and distribution of ongoing work via online preprint archives. Scholars conducting exploratory research over a collection of papers often exhibit opportunistic and dynamic information needs~\cite{soufan_searching_2022}, and managing papers across historical collections and continuous publication streams can be challenging~\cite{sultanum_understanding_2020,mysore_how_2023}. To support scholars in triaging, organizing, and reviewing papers, a plethora of systems have been introduced within both academic research and industry contexts.

One line of work aims to support scholars in exploring a collection of papers. For instance, many scholars use AI-powered systems such as paper search engines (e.g., Semantic Scholar~\cite{ammar_construction_2018} and Google Scholar~\cite{beel_google_2009}), graph-based visual tools (e.g., \textit{ConnectedPapers}~\cite{connectedPapers}), and information extraction tools (e.g., \textit{Elicit}~\cite{elicit}) to assist in discovering relevant work or scaffolding a literature review. Prior work has also developed research prototypes that facilitate visual exploration of the research landscape~\cite{ponsard_paperquest_2016, he_paperpoles_2019, matejka_paper_2021, wang_guided_2016, sultanum_understanding_2020, chau_apolo_2011}, scaffold paper exploration through augmentations of related work sections~\cite{palani_relatedly_2023} and citations~\cite{chang_citesee_2023, rachatasumrit_citeread_2022}, leverage personalized cues for paper recommendations~\cite{kang_who_2022} and literature discovery~\cite{kang_comlittee_2023}, and synthesize research threads across papers~\cite{kang_threddy_2022, kang_synergi_2023}.

Another set of interactive systems aims to support scholars in reading and understanding individual papers. For instance, \textit{PaperPlain} helps lay readers navigate medical literature with AI-generated section summaries and suggested questions~\cite{august_paperplain_2023}, and \textit{ScholarPhi} helps scholars understand paper-specific jargon with definition and equation augmentations~\cite{head_augmenting_2021}, allowing scholars to click on specific terms and view definitions in-situ within a floating card. We adopt a similar interaction design in \sys{}, enabling interactions on suggested entities underlined within an abstract to create expansions with minimal effort (i.e., a single click).

Other systems draw on visual aids to improve comprehension, for instance by highlighting conceptual relationships within papers using bubble-tree map visualizations~\cite{zhang_conceptscope_2021}, embedding animated figures into papers~\cite{grossman_your_2015}, and linking video summaries from authors' talk videos with relevant passages in a paper~\cite{kim_papeos_2023}. Complementing these tools that support paper comprehension, some systems facilitate rapidly reading or skimming papers. For instance, \textit{Spotlights} anchors visually salient objects as transparent overlays on a paper to facilitate high-speed skimming~\cite{lee_spotlights_2016}, and \textit{Scim} uses faceted highlights to direct readers' visual attention through a paper~\cite{fok_scim_2023}. Our work presents an interaction technique situated at the intersection of supporting paper exploration and comprehension. Specifically, we seek to narrow the informational gap between a paper's abstract and full text, addressing scholars' personalized information needs as they arise during the triage.

\section{Recursively Expandable Abstracts}

\begingroup

\renewcommand{\arraystretch}{1.2}

\begin{table*}[ht]
    \small
    \centering
    \Description[]{}
    \caption{Participants in our formative study and selected questions they asked while reading scientific paper abstracts.}
    \begin{tabular}{p{0.02\textwidth}p{0.28\textwidth}p{0.6\textwidth}}
    \toprule
        & \textbf{Title (Research Area)}      & \textbf{Representative Questions} \\
    \midrule
    P1  & PhD student (Explainable AI)  & What does ``TAM'' mean? What is a ``path analysis''? What does ``visual question answering task'' mean? What's an example of a ``trustworthiness cue''? \\
    P2  & PhD student (Computational Biology)  & What is the ``two-stage algorithm''? What are the 12 challenging reasoning tasks? What's an example of this task?  What is ``the zero hypothesis''? What does ``outlying failure cases'' mean? \\
    P3  & PhD student (Human-AI Interaction)  & What does ``comparable to SoTA phrase based systems'' mean? What is a ``sequence transduction model''? What is the ``path-x challenge''? What is the ``path-256 task''? What does ``retrieval set'' mean? \\
    P4  & PhD student (NLP)  & What loss function did they use? What’s the model architecture? What’s the metrics they use? Why do they call it ``human-interpretable''? \\
    P5  & Post-doctoral scholar (HCI) & How do they define ``trust'' in human-AI teamwork; is this reliance? Is ``human-agent teamwork'' a defined sub-field of prior work, or is that just jargon the authors use? What are examples of ``spatial crowdsourcing''? What are examples for the ``two realistic task assignment settings''? \\
    P6  & PhD student (HCI)  & What do the authors mean by ``gigification of knowledge work''? What were the design recommendations? What are the key findings in a simplified sense? What papers are they building upon? What is their study design?  \\
    P7  & PhD student (HCI)  & What does ``perceived valence'' mean? What does ``participatory foresight'' mean? What do they mean by ``early testing of AI-based features''? What does a report look like when they say ``practitioners with reports''? \\
    \bottomrule
    \end{tabular}
    \label{tab:formative-study-overview}
\end{table*}

\endgroup


\subsection{Formative Study}
To understand the types of information needs that emerge when exploring scientific paper abstracts, we first conducted a formative study observing scholars reading abstracts in a familiar domain. 

\subsubsection{Participants}
We recruited seven participants from an academic institution via social media and snowball sampling (Table~\ref{tab:formative-study-overview}). All participants actively conducted research across different areas of computer science, and reported familiarity with the research process, including triaging, reading, and organizing scientific papers. 

\subsubsection{Procedure}
After introducing the study and obtaining consent, participants were asked to read 3--5 abstracts of their choice. Participants selected abstracts from various sources: many used results from a paper search engine seeded with a relevant prompt for their area of research, one used abstracts for papers they were currently reviewing, and one used papers they had previously saved for a later reading session. While reading each abstract, participants were asked to share aloud any thoughts, questions, or confusions they had about the information within the abstract or about the paper in general. All studies lasted about 45 minutes. Participants were thanked for their time.

\subsubsection{Findings}
Our observations revealed four common types of information needs participants had while reading an abstract:

\paragraph{Definition} Participants wanted to define jargon, unfamiliar language, or acronyms they encountered in the abstract. The definitions they sought were often not in the general sense, but rather specific to the context of the paper they were currently reading.

\paragraph{Instantiation} Participants sought examples to provide concrete context for under-specified language. For instance, in the sentence, ``We find our approach outperforms three baselines on a common question-answering benchmark,'' it is unclear which specific baselines or question-answering benchmark was used. When an abstract described an unfamiliar concept, e.g., a new task or dataset, participants also wanted to view an instance of the concept to help visualize its structure and compare it against familiar concepts.

\paragraph{Clarification} Participants sought additional context to help explain technical or unfamiliar language in an abstract. Since abstracts are concise, self-contained summaries for a long paper, authors are compelled to withhold particular details and use dense language to convey information. As a result, scholars reading abstracts often have information needs expressed through implicit clarification questions, personalized by their own expertise and reading goals.

\paragraph{Motivation} Participants expressed a desire to probe the authors’ motivations and justifications for aspects of the paper. For instance, some participants asked why the authors chose their particular method (e.g., model architecture, loss function, or task), why particular language was used to describe their system (e.g., ``human-interpretable''), or the signifance and novelty of their research problem. Addressing this need could help satisfy curiosities, expand their understanding, or evaluate a paper's validity.


\subsection{Recursively Expandable Abstracts Interaction and Design Space} \label{sec:designSpace}
\begin{figure}[t]
    \centering
    \includegraphics[width=0.47\textwidth]{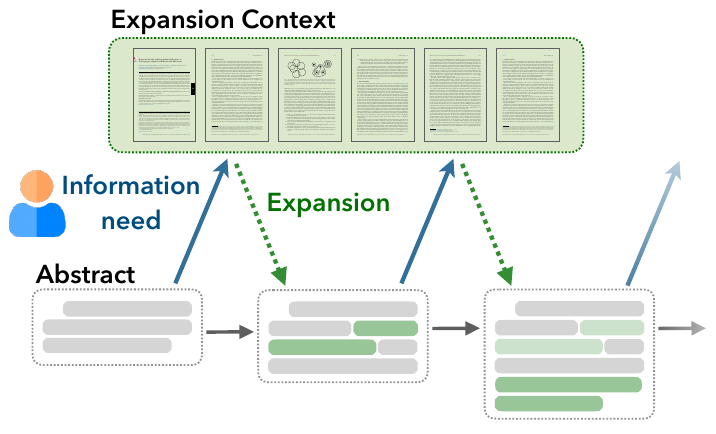}
    \caption{Recursively expandable abstracts allows users to retrieve clarifying information from a broader \textit{expansion context} (e.g., the full paper) in response to dynamic information needs, forming \textit{expansions} that grow the abstract fluidly.}
    \Description[]{}
    \label{fig:abstract-expansion}
\end{figure}

Based on the information-seeking behaviors found in our formative study, we propose an interaction technique called \textit{recursively expandable abstracts}, that augments abstracts with additional relevant information in response to dynamic user queries for clarification (Figure~\ref{fig:abstract-expansion}). This expandable text paradigm is inspired by StretchText (or transclusion)~\cite{bernstein_hypertext_2009, nelson_literary_1983}, an early vision in Project Xanadu and hypertext design from the 1980s for structuring text on the web that allows users to choose the level of detail they want to see. When a specific area or keyword is selected, the originally concise text ``stretches'' to reveal additional details. While the original vision for StretchText requires carefully-authored, structured text and has not gained wide adoption, we revisit and build on this vision by leveraging LLMs to dynamically generate on-demand summaries that support personalized and interactive expandable text.

\begin{figure}[t]
    \centering
    \includegraphics[width=0.46\textwidth]{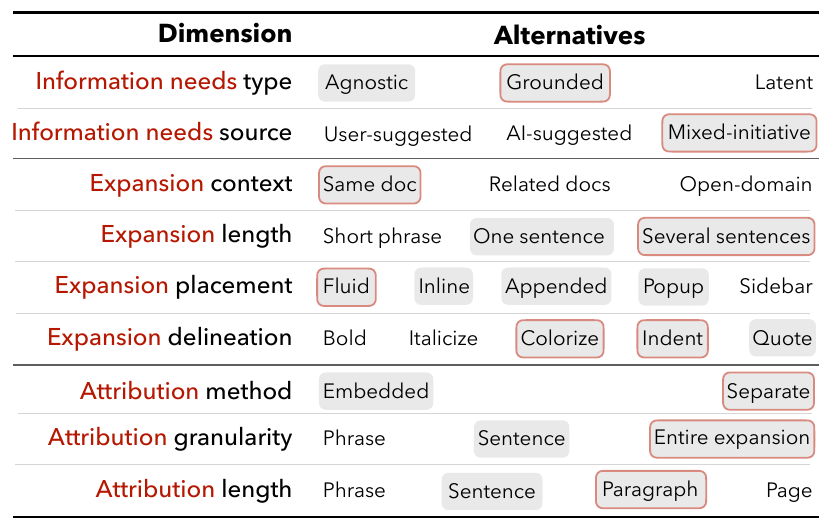}
    \caption{The design space for an expandable abstracts interaction paradigm. Alternatives we explored in \sys{} are highlighted in gray, and those included in the final system are outlined in red.}
    \Description[]{}
    \label{fig:design-space}
\end{figure}

For this, we focused on designing expandable summaries based on scientific abstracts and papers. Abstracts are concise, static, author-crafted text summaries for a long scientific document; as such, no single abstract can concisely capture the interests of every reader or address dynamic information needs that arise while reading \citep{zhang_concepteva_2023}. Expandable abstracts ameliorates the static limitations of summaries by allowing scholars to interactively expand abstracts with additional clarifying information retrieved from an \textit{expansion context}. The expansion context for many clarification questions arising from an abstract is often the corresponding paper's full text. However, expansions could also be drawn from information in a broader domain, such as related papers in the paper's citation network or general information in an online resource (e.g., Wikipedia).
To inform the interaction design, we first articulate the plausible dimensions and alternatives of its design space (Figure~\ref{fig:design-space}).

\subsubsection{Information needs}
One set of dimensions of the design space concerns what and how information is selected for expansion. First, what \textit{type} of information needs should be expanded? The four types of needs identified in the formative study are similar in that they all represent an information need \textit{grounded} in language from the abstract. These may emerge as scholars read an abstract, anchored to specific words in the abstract. For instance, in the sentence, ``We evaluated against three baseline approaches on a popular question-answering benchmark,'' a grounded information need might be: ``What were the three baseline approaches'' or ``What was the popular QA benchmark used?'' Information needs may also extend beyond the content explicitly stated in an abstract. Scholars might approach abstracts with predefined questions \textit{agnostic} to any particular abstract but relevant to their broader research goals. For example, they may seek to understand the methods, experiments, or findings across multiple papers as part of their exploration. Finally, \textit{latent} information needs refer to details that scholars may not consciously recognize, but are relevant to their goals. These details, although valuable for expansion, are not explicitly mentioned in the abstract or at the forefront of the scholar's consciousness, making them difficult to query for directly. Expansions could also be instantiated from different sources, for instance stemming from a user's question, an AI's suggested entity for expansion, or some mixed-initiative approach.

\subsubsection{Expansion}
A second set of dimensions considers the content and visualization of information within an expansion. Information used to expand an abstract could be retrieved from different \textit{contexts}, e.g., the full text of the paper for an abstract, other relevant papers, or an external knowledge base such as Wikipedia. The \textit{length} of the generated expansion is also an important consideration, with good designs aiming to balance addressing a scholars' information needs yet being judicious in length to limit the added cognitive load. Generated expansions may range from a phrase or sentence, to a longer paragraph with details that could motivate further exploration.

The \textit{placement} of expansions within a system interface is another important design choice. Possibilities include placing expansions within an adjacent pane (similar to many chat-based applications with documents), in a popup card (similar to citation cards in augmented paper reading interfaces or page previews in Wikipedia), appended at the end of a summary, or placed in-situ near an appropriate text anchor in the abstract. Selecting the optimal placement requires trading off the navigational effort between an expansion and the abstract, and the potential for visual distraction or clutter. For placements that interweave expansions and the original abstract, it is important to consider how a system may differentiate between these two sources of information. Unlike the original abstract, whose provenance is known and trusted, expansions are AI-generated, introducing concerns around hallucination or trustworthiness of the generated information. Approaches may include visually delineating between the expansion and abstract text with standard visual cues, e.g., bold, italics, color, or through positional displacement, helping users to visually identify and switch between the two sources of text.

\begin{figure*}[t]
    \centering
    \includegraphics[width=0.95\textwidth]{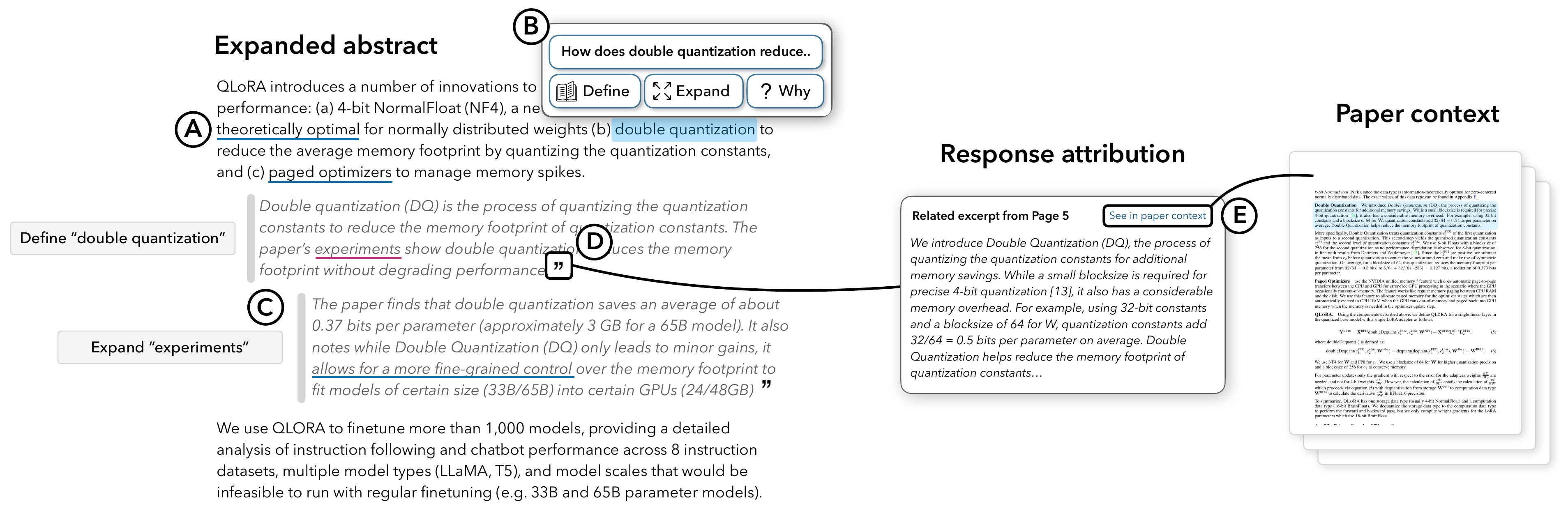}
    \caption{Recursively expandable paper abstracts with attribution in \sys. Expansions are created on-demand by highlighting text in the abstract or selecting an AI-suggested expandable entity (A), revealing a question palette (B). Selecting a question in the palette prompts an LLM to retrieve relevant clarifying information, presented as a fluid expansion within the abstract (C). Users can drill-down to see evidence for a response in a paper excerpt (D) and within the full paper context itself (E).}
    \Description[]{}
    \label{fig:ui_overview}
\end{figure*}


\subsubsection{Attribution}
A third set of dimensions considers designing to convey information provenance for the generated expansions. Provenance for question-answering contexts is is often achieved through attribution, i.e., retrieving evidence from the expansion context that support the generated answer. There are several \textit{methods} of presenting attribution---evidence could be embedded directly in an expansion, akin to quotes, and use visual cues to demarcate abstractive and extractive text, or provided as separate excerpts surfaced verbatim from the expansion context.
For longer or more complex expansions, multiple pieces of evidence may be necessary to support each claim in the generated text. In these cases, systems may consider the most appropriate \textit{granularity} of attribution to determine the utility of providing evidence for each phrase, sentence, or entire expansion. Finally, selecting the appropriate \textit{length} of attributed evidence can ensure sufficient validation of the generated expansion without introducing excessive cognitive burden.

\section{The \sys{} System}
Next, we describe the design and implementation of \sys{}, an augmented reading interface that implements recursively expandable abstracts. \sys{} affords interactive expansion of abstracts on-demand, progressively incorporating information from the full paper relevant to users' dynamic goals and information needs.

\subsection{User Interface} \label{sec:user_interface}
The design of \sys{} was guided by the four types of information needs observed in our formative study, and refined through an iterative design process in which alternatives of the design space were implemented and evaluated.

\subsubsection{Eliciting Information Needs as Clarifying Questions}
With \sys{}, users begin by reading an abstract as they typically would. As information needs arise during reading, users can highlight any span of text within the abstract to request additional information. In response to a user's highlight, \sys{} displays a \textit{question palette} centered above the highlighted text (Figure~\ref{fig:ui_overview}B) that enables users to easily specify their information needs as clarifying questions anchored to a specific context in the abstract (i.e., the highlighted text). The question palette contains four buttons: three static buttons with the questions \textit{Define}, \textit{Expand}, and \textit{Why}, and one dynamic button with an AI-suggested question.

The AI-suggested question aims to predict a user's intent, offering the most probable clarification question a user may want to ask given the text they selected. The three static questions are fixed regardless of the selected text, reflecting the common grounded information needs revealed in our formative study. Specifically, the \textit{Define} question aims to address \textit{Definition} and \textit{Instantiation} needs, the \textit{Expand} question aims to address \textit{Clarification} and \textit{Instantiation} needs, and the \textit{Why} question aims to address \textit{Motivation} needs. The \textit{Expand} question is visually centralized, serving as a ``catch-all'' option to incrementally retrieve more details appropriate for most information-seeking circumstances.

While increasing the number of static questions could provide more flexibility, based on feedback to initial prototypes of \sys{} we determined that providing more questions could clutter the interface, occlude more of the abstract, and cause decision paralysis in selecting an appropriate question. For similar reasons, only the top-1 AI-suggested question is shown in the question palette.

To complement the manual highlighting of text users want to expand, \sys{} also pre-selects several \textit{expandable entities}. These entities capture spans of text in the abstract that the system believes could be further expanded from the expansion context; for instance, they could include under-specified language (e.g., ``some'', ``several'', ``various'') or jargon (e.g., acronyms). Expandable entities are visually indicated with a blue underline (Figure~\ref{fig:ui_overview}A), and users can click on an entity to reveal the question palette. Altogether, \sys{} aims to reduce the cost of asking grounded information-seeking questions through these two lightweight interactions.

\subsubsection{Expanding Abstracts with Clarifying Information} 
When users select a question from the question palette, \sys{} creates an \textit{expansion} by fluidly expanding the abstract with in-situ information retrieved from a larger expansion context (Figure~\ref{fig:ui_overview}C). Each expansion is an abstractive, LLM-generated response to a user's question, containing up to three sentences. In instances where a question cannot be answered, no expansion is created, and a toast alert is shown in the bottom right of the screen instead to indicate an expansion could not be created.

Key to the expansion's design is ensuring visual delineation between the original text of the abstract and that of the generated expansion. Expansions are presented as indented blocks, appended below the sentence containing the selected expandable entity. A vertical bar and tag containing the question used to generate the expansion are shown to the left of each expansion to help users easily identify the visual boundary and purpose of each expansion. The text of an expansion is also colored blue when initially created, gradually transitioning into a light gray after a few seconds. This produces a smooth animation that visually cues attention to a new expansion, before fading into a color (gray) that is similar but not overly distracting from the color of the original abstract text (black). To indicate parts of the abstract previously expanded, text used to seed the expansion---either from a user highlight or an expandable entity---is underlined in purple.

For each generated expansion, \sys{} suggests additional entities that could be expanded. Similar to interactions with the original abstract text, users can select an AI-suggested entity or highlight any text in the new expansion to recursively expand further. Expansions created on other expansions form a threaded abstract reading experience, allowing users to easily dive deeper into aspects of interest from the abstract by gradually retrieving details from the full paper. As each expansion is limited to three sentences in length, users are less likely to be overwhelmed by information in any single expansion. They can continue expanding until their information needs are satisfied, at which point they can easily pop back up into the original abstract or expansions at any level. If a particular expansion is no longer needed, users can click on the question tag at the left of the expansion to collapse it into its parent expansion.

\subsubsection{Drilling Into a Paper with Attributed Responses}
Since each expansion is generated automatically by an LLM, there is a potential risk of generating content that is unfaithful to the original paper or otherwise factually incorrect, a problem referred to as hallucination~\cite{maynez_faithfulness_2020, ji_survey_2023}. To help mitigate these risks, \sys{} provides \textit{attribution}, or extractive supporting evidence, for each expansion. Users can click on a quote button at the end of each expansion to show a card with the most relevant paragraph from the full paper (Figure~\ref{fig:ui_overview}D). Within the card, users can further drill-down to open the paper in an integrated document viewer overlay, with the attributed paragraph navigated to and highlighted in the context of the paper (Figure~\ref{fig:ui_overview}E). Through these two levels of interactive attribution, \sys{} allows users to explore the surrounding paper context and verify the accuracy of an expansion.

\subsection{System Architecture}

\begin{figure*}[t]
    \centering
    \includegraphics[width=0.95\textwidth]{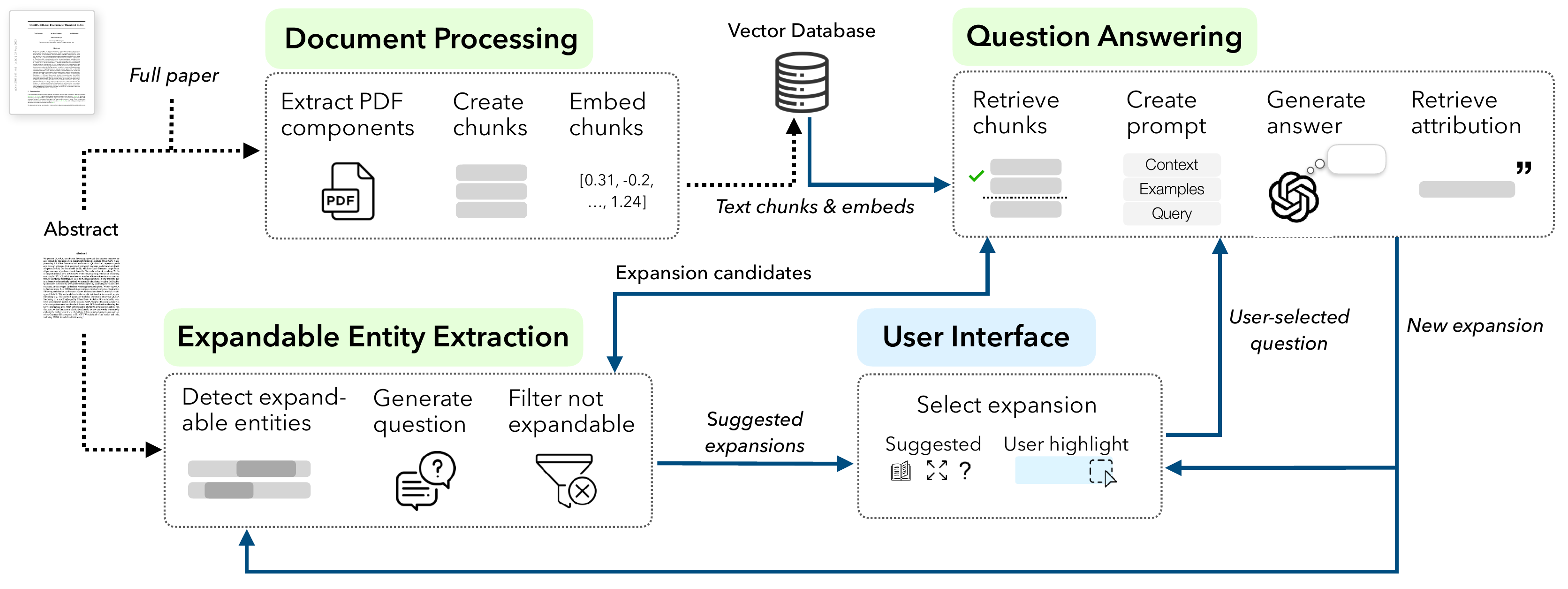}
    \caption{\sys's system architecture. Each paper is first processed (Document Processing) and initial expandable entities are extracted from the abstract (Expandable Entity Extraction). When a user asks a question for an expandable entity, \sys{} uses a retrieval-augmented generation approach to form a response and retrieve attribution (Question Answering). \sys{} then suggests expandable entities within the response, allowing recursive expansions.}
    \Description[]{}
    \label{fig:system_pipeline}
\end{figure*}

To create an expandable abstract, \sys{} implements three natural language services (illustrated in Figure~\ref{fig:system_pipeline}): (1) A \textit{document processing} service extracts and encodes information for a paper's full text; (2) A \textit{question answering} service generates attributed answers to users' document-centered questions; and (3) An \textit{expandable entity extraction} service identifies expansion candidates within an abstract or generated expansion. We provide an overview of our implementation of these services, which may serve as a starting reference for others exploring similar expandable summary interactions.

\subsubsection{Document Processing}
Papers ingested by \sys{} are first processed to reduce latency when interactively generating expansions at query-time. Each PDF is parsed into its constituent tokens and bounding boxes, and sentences and paragraphs are constructed from the full text. Then, chunks are created with a chunk size of three sentences and a two sentence overlap. Each chunk is embedded using the \texttt{all-mpnet-base-v2} encoder from the SentenceTransformers framework~\cite{reimers-2019-sentence-bert} and stored into a vector database. Embeddings of paragraphs are also created and stored in a separate index, which are used for retrieving attribution.

\subsubsection{Attributed Question Answering} \label{sec:attributed-qa}
We use a retriever-reader architecture with LLMs for retrieval-augmented question answering. When users select a question from the question palette, \sys{} first embeds the question with the same encoder used during processing. It then retrieves the 12 most relevant paper chunks (with relevance determined by cosine similarity between chunk and question embeddings) to form a context. An LLM prompt is then formed by concatenating a general description of the question answering task, the context, a few examples of question and answer pairs, and the question. The prompt further instructs the LLM to be concise, use language from the provided paper context when appropriate, generate answers containing no more than three sentences, and return no answer if the question cannot be answered given the context. These specific prompt tuning adjustments were made over several iterations of inspecting \sys{}'s expansions. Finally, we use \texttt{gpt-3.5-turbo-1106} to generate an answer for the question using this few-shot prompt. The current \sys{} prototype answers questions using information from the full text of the source paper only; we leave consideration of other possible expansion contexts (e.g., other related papers) for future exploration.

\sys{} further provides attribution for each of its expansions to enable users to verify the accuracy of the generated answer and ease into the full paper. To generate attributions, \sys{} retrieves the most relevant paragraph to the generated text (by cosine similarity). We explored other attribution schemes in earlier iterations of \sys{}. For instance, we tried retrieving chunks for each individual sentence, but found chunks were less preferred than paragraphs since they sometimes lacked sufficient context. We also tried providing attribution for each sentence where expansions consisted of multiple sentences. However, we found the need to read and reconcile multiple attribution sources introduced confusion and made verification more challenging.

\subsubsection{Expandable Entity Extraction}
To complement users in manually specifying their own expansions, \sys{} proactively suggests parts of an abstract or expansion that could benefit from additional context. To identify these regions within an abstract, \sys{} uses \texttt{gpt-4-1106-preview} with a few-shot prompting strategy. The model is instructed to identify short text spans (i.e., entities) which may be expanded to provide clarification for vague, dense, or jargon-rich language. The prompt also specifies that information required to expand each entity should not be already available in the abstract or expansion. For each entity, \sys{} performs a dry-run expansion (using the same \textit{Expand} question in the question palette), and removes entities for which no answer is found.

For each remaining entity, \texttt{gpt-4-1106-preview} is directed with zero-shot prompt to generate a single question that users might ask in expanding that entity. For instance, given the sentence, ``We propose a new framework to address the ACTA task,'' \sys{} could identify the entities ``a new framework'' and ``ACTA,`` and generate the questions ``What are the main characteristics of the proposed framework?'' and ``What is the ACTA task?,'' respectively. All of these \textit{expandable entities} are underlined in reading interface and the suggested question is shown in the question palette. The same question generation prompt is executed on-the-fly to generate the suggested question within the question palette when users create an expansion by highlighting any text.

\subsection{Implementation Details}
\sys{} was implemented as a standalone web application using TypeScript, CSS, and the React framework~\cite{react} for the user interface. Backend services and LLM-powered functions were implemented in Python and the Flask framework~\cite{flask}. GROBID~\cite{grobid} was used to parse paper PDFs into a structured JSON format, and the MMDA open-source library~\cite{mmda} was used to construct \texttt{Document} objects from the output, from which full text paragraphs and sentences could be retrieved. The PDF reader for viewing expansion attribution in context was adapted from an open-source PDF reader library~\cite{pdf-component-library}. The \texttt{gpt-3.5-turbo-1106} (with 16k token context window) and \texttt{gpt-4-1106-preview} (with 128k token context window)\footnote{At the time of submission.} LLMs were accessed via OpenAI's APIs with a generation temperature of 0 and maximum length of 750 tokens. The specific LLM prompts we used for each service is provided in Appendix~\ref{appendix:llm-prompts}. 


\section{Study 1: Interview Study}
To evaluate the usability and utility of \sys{}, we first conducted a qualitative interview study using a think-aloud protocol. In this study, we aimed to answer two research questions:

\begin{enumerate}[label=\textit{RQ\arabic*.}]
    \item How do users engage with \sys{} for exploring abstracts?
    \item What are the advantages and limitations of AI-augmented abstracts?
\end{enumerate}

\subsection{Study Design}

\subsubsection{Participants}
We recruited 9 researchers (6 male, 3 female; Age: M = 27.8, SD = 9.3) via university mailing lists and Slack channels. Eight participants were doctoral students within computer and information science and one participant was a research scientist.

\subsubsection{Selected Papers}
To incentivize engagement, for each participant we curated a personalized set of abstracts that aligned with their research expertise and interests. In a screening survey, we asked participants to list up to 5 ``seed'' papers representative of their research interests. We then used the Semantic Scholar Recommendations API\footnote{https://api.semanticscholar.org/api-docs/recommendations} to obtain 25 additional recommended papers for each participant based on their seed papers. We created expandable abstracts within \sys for all seed and recommended papers, excluding those without a valid PDF file.

\subsubsection{Procedure}
Participants first completed a tutorial that introduced them to \sys{}'s features (5 min). They were then instructed to browse their personalized list of expandable abstracts, imagining the list was recommended to them from a colleague or search engine (25 min). During their exploration, participants were asked to think aloud, sharing any observations or questions as they emerged and explaining the interactions they chose to use. After the task, participants were engaged in semi-structured interviews that sought to elaborate on the advantages and limitations of \sys's expandable abstracts (15 min). All think-aloud sessions and interviews were recorded, transcribed, and coded through a thematic analysis (additional details in Appendix~\ref{appendix:interview-questions}). Study sessions lasted 45 minutes and participants were compensated with \$25 USD. This and all following studies involving human subjects were approved by a university institutional review board.

\subsection{Results}
An analysis of interview transcripts and interaction logs uncovered various ways in which \sys{} supported the exploration of abstracts, such as using interactive expansions to retrieve additional information on-demand from full papers, threaded exploration to dive deeper into aspects of interest, and LLM-generated expandable entities and questions to guide attention. In the following results, we refer to participants with the pseudonyms P1--9.

\subsubsection{Abstract expansions allowed an on-demand recursive extraction of high-quality information from papers}
Participants were actively engaged with \sys{} during the study. On average, each participant explored 4.0 papers (\textit{SD} = 1.1, \textit{Mdn} = 4.0) and created a total of 20.8 expansions (\textit{SD} = 7.8, \textit{Mdn} = 18.0). Based on the think-aloud, participants liked how \sys{} allowed them to surface details from the paper using simple interactions with the abstracts over manually searching for them over the full papers.
For example, one participant remarked, \textit{``I was impressed by the things that I was able to pull from the paper and the amount of additional details I can get just by reading this abstract''} (P3). Participants also pointed to how abstracts have a familiar structure that served as natural entry-points to scaffold pulling in additional details from parts of the full paper when needed:
\begin{quote}
    \textit{``I think that one of the key things about being a PhD student is being able to quickly break down information without having to spend too much time reading the entirety of the paper. And so the abstract allowed for me to do that without having to even read that. Usually the rule of thumb is abstract, intro and conclusion. And with this, I feel I get a bit of the intro, conclusion, results, discussion, analysis, all that within the abstract breakdown.''} -- P7
\end{quote}

Participants further noted how \sys{}'s generated expansions answered the questions they asked surprisingly well (P1--3, P5, P6, P8). One participant appreciated how the expansions \textit{``didn't just summarize, but seemed to extract meaning from the paper''} (P6), validating the LLM's ability to form concise and useful answers drawn from complex text. We observed some participants beginning their exploration by browsing the abstract of a paper they were familiar with, attempting to gauge the accuracy and reliability of the generated expansions (P5, P8, P9). Others instead mentioned trusting the model's responses, especially with the confidence that they could dive into the paper to verify the attributed evidence if needed (P4, P7, P8).

Participants found the AI-suggested question in the question palette often aligned well with their information-seeking intents and reduced the costs of forming a question to expand the abstract. P8 described the suggested question as, \textit{``It seems to almost read my mind when I click on something or highlight something.''} 
,and P5 remarked, \textit{``Every time I think of what the question is, that's pretty much what the question it already thought of is.''} A similar sentiment was reflected in \sys{}'s usage behaviors; within the question palette, participants selected the LLM-generated question most often (40.1\% of clicks, \textit{Define}: 23.5\%, \textit{Expand}: 22.5\%, \textit{Why}: 13.9\%).

Participants also frequently utilized the recursive expansion feature of \sys{}---58\% of expansions created by participants were threaded (i.e., formed by asking questions about text in another expansion rather than from the abstract). Based on their think-aloud, participants mentioned how the threaded expansions empowered them to dig deeper and ask follow-up questions if an initial expansion did not fully satisfy their information needs (P1, P3, P4, P6, P8). 
For example, P4 said, \textit{``I really did love the way you could keep going and go branching into a tree.''}
Similarly, P1 described how the continuous presence of underlined AI-suggested entities served to motivate further exploration:
\begin{quote}
    \textit{``I can keep diving more because the highlighting feature is not provided only on the first level of the abstract, it’s also provided on the next level. When its generates a response in the second level, it also provides these underlining features, which if I have any more questions I could keep using these interactions to help me make sense of the abstract.''} -- P1
\end{quote}
In summary, the threaded expansions not only supported ongoing engagement but also sensemaking efficiency within the abstract.

\subsubsection{AI-suggested expandable entities guided and motivated deeper exploration of abstracts.}
The majority of the time, participants clicked on a pre-selected underlined entity to create expansions (77.5\%) as opposed to manually selecting a custom text span (22.5\%). While we initially designed this feature to lower the interaction costs, the think-aloud suggested that the pre-selected entities also served as information cues that can facilitate discovery.
Many participants (P1--4, P6, P8) commented how the underlined entities served as visual cues to \textit{``keywords that may be relevant''} and \textit{``tell me what to focus on''} (P1).

Conversely, if the underlines did not precisely capture participants' needs, the ability to highlight custom text spans to create an expansion was appreciated (P1, P2, P8). P6 summarized how the interactive features of \sys{} in concert could address all of her envisioned information needs over an abstract:
\begin{quote}
    \textit{``I think the underlines were already really good. Highlighting something as a backup to the underlines already gets you to probably 99\% coverage for the things you would want to ask. And then for the last 1\%, I could click on the quote and go into the paper.''} -- P6
\end{quote}

On the other hand, visually augmented reading interfaces such as \sys{} and its underlined entities can introduce distractions for some scholars.
Since expanding the abstracts by interacting with the underlined entities required less effort than highlighting text, P4 noticed how she \textit{``tended to default to whatever was already underlined,''} and it became \textit{``hard to remember that I can just like pick anything out unless I was really curious about it.''} This behavior is not necessarily undesirable, but suggests careful consideration should be given to how augmented interface elements may inadvertently guide or constrain user interactions.
Some suggested how the underlined entities could open up a rabbit-hole of exploration, derailing the reading of an abstract (P3, P7)
P7 further noted how the seemingly limitless freedoms afforded by an abstract expansion interaction could be double-edged and inhibit a sense of completion:
\begin{quote}
    \textit{``The endlessness of the underlines, as a completionist, my mindset, I want to click them all. And so I liked the fact that there were no bounds, but I could also feel overwhelmed knowing there are no bounds. So I feel like I could miss something the AI could uncover for me if I just kept clicking all the underlines.''} -- P7
\end{quote}
These observations suggest that while the underlines may help guide an in-depth investigation of details in the paper, it can also potentially hinder the process of triage. As such, balancing interactivity and efficiency is critical for an expandable abstract interaction, ensuring users can engage deeply when needed but also efficiently navigate the content within the abstract and generated expansions.

\section{Study 2: Field Deployment Study}
To further investigate how scholars would interact with expandable abstracts in the wild, we deployed \sys{} during the 49th International Conference on Very Large Data Bases (VLDB 2023). For the duration of the conference, members of the research team invited conference attendees and other scholars to try out \sys{} via social media announcements including email, Twitter, Slack, and LinkedIn. We created a landing page within \sys{} to allow users to easily browse the 248 papers within the conference proceedings, with a paginated, scrollable list of paper metadata and a search bar for filtering papers. Clicking on a paper title navigated to a separate page with an expandable abstract for that paper.

During the week of the conference and two subsequent weeks, a total of 275 unique users interacted with 50 unique papers using \sys{}. Based on the interaction logs, each user created 3.3 expansions on average (\textit{SD} = 4.6, \textit{Mdn} = 2.0). We found users expanded abstracts using the pre-selected expandable entities rather than manually selecting custom text spans (80.4\% vs 19.6\% of interactions), similar to findings from the interview study.

We further found that users more often selected one of the three static questions (i.e., \textit{Expand}, \textit{Define}, and \textit{Why}; 88.2\% of interactions) than the more specific LLM-generated questions
Across the static questions, users created 41.6\% of expansions with \textit{Expand}, 31.5\% with \textit{Define}, and 15.1\% with \textit{Why}, a distribution that closely corroborates the frequency of question types we observed in our formative study. These results are in contrast to our interview study, where participants were more likely to select LLM-generated questions (40.1\% of expansions). This difference might be due to the higher relevance of abstracts in our interview study compared to our deployment (i.e., paper abstracts were selected based on personalized recommendations for each participant). 

While the majority of the time users created a single level of expansion from the original abstract, a significant portion (27.7\%) of interactions were threaded, meaning users recursively asked follow on questions by selecting additional entities in the expanded text. Some users recursively created up to 5 nested expansions. These results demonstrate the users' needs for recursive expansion of abstracts and \sys{}'s ability to support this. In comparison to our interview study, fewer users in our deployment study created threaded expansions (58.0\% vs. 27.7\%, respectively), perhaps due to differences in user engagement or relevance of abstracts.

We also observed users actively engaging with the attributed evidence paragraphs and viewing the paper itself.
In 14.8\% of expansions, users viewed the attributed evidence paragraph for a generated expansion. About 60\% of the time users were satisfied with the extracted evidence, while 40\% of the time they further opened the PDF to view the highlighted evidence in the context of the paper. These behaviors suggest users were either interested in understanding the supporting information or sought to use the evidence as an efficient entry point into reading the paper.


Altogether, our deployment study suggests that an expandable abstract interaction presents a simple yet effective means to elicit clarification questions for abstract-grounded information needs, allowing users to retrieve attributed answers on-demand. Our findings complement the interview study and characterize real world usage behaviors of \sys{} within a natural context of user interaction where scholars are actively triaging papers.

\section{Study 3: Comparative Evaluation}
Finally, to understand how \sys{} compares to existing modes of triage, we conducted a within-subjects study where we compared \sys{} to two other paper exploration strategies. In this study, we aimed to answer the following research questions:
\begin{enumerate}[label=\textit{RQ\arabic*.}]
    \setcounter{enumi}{2}
    \item How does \sys{} affect the quantity and types of questions users ask over abstracts and papers?
    \item How does \sys{} compare to and affect users' current navigational strategies across abstracts and papers?
\end{enumerate}

Based on feedback from the previous interview and field deployment studies,  we made small refinements to the design and usability of \sys{}.  Specifically, we enabled users to ask any question they desired by editing the AI-suggested question in the question palette. We also removed the \textit{Why} question as it was the least frequently used, and the additional option for expansion added to users' cognitive load, especially given that users could now compose their own questions. Lastly, we moved the quote button for displaying attributed evidence from inline with the expansion text to the right margin adjacent to the expansion, and added a second button that hid a generated expansion, allowing users to easily view just the original abstract.

\begin{figure*}[ht]
    \centering
    \includegraphics[width=0.95\textwidth]{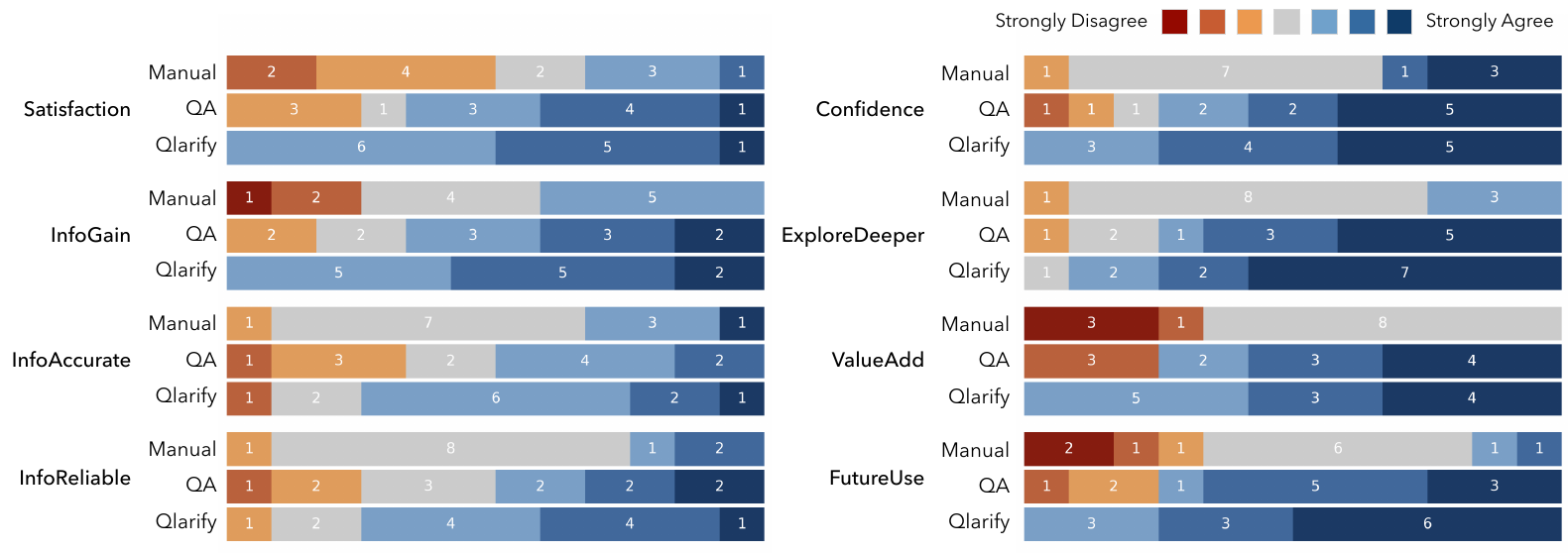}
    \caption{Distribution of participants' self-reported ratings within each condition in the comparative evaluation. Participants in the Qlarify condition felt more satisfied with their exploration, more confident in retrieving relevant information from the full paper, more motivated to explore deeply, and a greater desire to use in the future. See  Appendix~\ref{sec:comp_eval_survey_questions} for the precise wording used in the survey questions.}
    \Description[]{}
    \label{fig:survey_likert_responses}
\end{figure*}

\subsection{Study Design}
\subsubsection{Participants}
We recruited 12 researchers (9 female, 2 male, 1 non-binary; Age: \textit{M}=25.5, \textit{SD}=3.1), who had previously read at least one HCI research paper, via Slack and snowball sampling. 10 were doctoral students, 1 was a Master's student, and 1 was an undergraduate student. The doctoral students reported having between 1--5 years of research experience. 7 of 12 participants identified their primary discipline as HCI or related sub-fields (e.g., human-centered AI, tech policy), 2 as AI/ML, and 2 as robotics. All participants reported reading research abstracts and papers at least once a week; a majority reported reading more than 10 abstracts and 1--2 full papers on average each week.

\subsubsection{Conditions}
The study included three system interface conditions: Manual, QA, and Qlarify. A comparison of the interfaces is provided in Figure~\ref{fig:study3_conditions} in the Appendix. In the Manual condition, participants were given a list of paper titles and other metadata (authors, venue, publication year), abstracts, and PDFs, reflecting the manual process of browsing papers and abstracts. In the QA condition, participants were given the same paper elements as in Manual, but with an additional text field, below the paper abstract, that allowed users to sequentially ask questions about the paper. Responses to submitted questions were generated by an LLM (using the retrieval-augmented generation method in \S\ref{sec:attributed-qa}) and placed in a box below the question text field. The QA condition aimed to emulate a simple question answering service over full papers, and did not share context between multiple questions or provide attribution. In the Qlarify condition, participants were given the same paper elements as in Manual, but additionally had access to all features of \sys{} as described in \S\ref{sec:user_interface}.

\subsubsection{Procedure}
Participants were first provided with a tutorial introducing them to the three conditions (15 min).
They then completed a task of reviewing papers from UIST 2023, selected based on the availability of PDFs (80 out of 123 papers). They were asked to envision themselves as TAs for an HCI seminar class and to create a shortlist of papers suitable for student reading, discussion, and presentation. This collection was selected to ensure the participants had relevant interest and expertise for triaging the papers. Participants completed the task using each of the three conditions (15 min each), the order of which was counterbalanced to mitigate the influence of any ordering effects. After the task, participants completed a post-task survey and a short interview describing their experience using the three systems. Study sessions were conducted remotely through an online video conferencing software and lasted 75 minutes. Participants were compensated with \$35 USD.

\subsubsection{Measures}

For quantitative data, we analyzed responses to the post-task survey in which participants rated their agreement with eight statements on seven-point Likert scale for each condition. These statements included participants' self-reported measures of satisfaction and confidence completing the task, the quality and reliance of gained information, and desire for future system use. Detailed survey questions are provided in Appendix~\ref{sec:comp_eval_survey_questions}. We also analyzed participants' interaction logs to measure the quantity and types of questions users asked and their patterns of navigation throughout the abstracts and papers (e.g., opening a paper's PDF). We used Wilcoxon signed-rank tests for two-condition comparisons and Friedman tests with Nemenyi post-hoc tests for three-condition comparisons. For qualitative data, we transcribed the post-task interviews and coded them through a thematic analysis. In this section, we refer to participants with the pseudonyms P1--12.

\subsection{Results}


\subsubsection{Asking Questions (RQ3)}
On average, participants asked more questions in the Qlarify condition than in the QA condition (Qlarify: 15.0 (\textit{SD}=5.4), \textit{Mdn}=15.5 (\textit{IQR}=11.3-19.0), QA: 9.6 (\textit{SD}=3.9), \textit{Mdn}=10.0 (\textit{IQR}=6.8-12.5), \textit{W}=4.0, \textit{p}=.02). 
Participants interacted with a similar number of abstracts in both conditions (Qlarify: 5.0 (\textit{SD}=1.9), QA: 4.7 (\textit{SD}=2.0)), but on average asked more questions per abstract using \sys{} (3.7 (\textit{SD}=2.9) vs. 2.2 (\textit{SD}=0.7)).
These findings suggest that \sys{} motivated participants to ask more questions about the abstracts than unaided question answering, with participants describing how the underlined expandable entities and lightweight interactions in \sys{} helped to reduce friction in formulating and asking questions about papers.
Specifically, we found that when using \sys{}, participants created an expansion by selecting an AI-suggested expandable entity 72\% of the time (as opposed to manually highlighting an entity 28\% of the time).
Within the expansion palette, participants selected the LLM-generated question most often (38.9\% of the time), in line with observations from the interview study. 
They also opted to use the newly added feature of editing the LLM-generated question with their own question within the question palette 11.1\% of the time, and used the two static questions approximately equally (\textit{Expand}: 27.2\% of clicks, \textit{Define}: 22.8\%).  
Many of the questions that participants asked were further threaded, i.e., not on the original abstract (30.3\%), suggesting the \sys{} encouraged participants to ask follow-up questions and dive deeper into specific parts of the paper through the abstract.

One disadvantage participants noted in the Qlarify condition was that the requirement of grounding in the abstract for generating expansions made it harder to ask general questions about the paper. Although possible by selecting a span and overwriting the AI-suggested question, this process felt less natural than asking questions in a blank text box. For P5, Qlarify and QA met different needs, saying that QA ``allows overall questions I had about the paper'' while ``Qlarify is useful for a deeper dive, aiding me in better understanding particular parts of the paper.'' Other participants used the QA condition to emulate familiar LLM-based chat interfaces; for example, one prompted multiple abstracts with, ``Explain this paper to me like I'm a 10 year old'' (P1). Consequently, many participants mentioned an ideal system would combine both conditions, offering interactive expansions in the abstract for grounded questions and a text box for general questions.


\subsubsection{Comparison with Current Navigation Strategies (RQ4)}
Overall, participants reported feeling more satisfied with their exploration of abstracts and papers in the Qlarify condition (6.2 (\textit{SD}=0.8), \textit{Mdn}=6.0 (\textit{IQR}=5.8-7.0)) than the Manual (4.8 (\textit{SD}=1.5), \textit{Mdn}=4.0 (\textit{IQR}=4.0-6.25)) and QA (5.5 (\textit{SD}=1.7), \textit{Mdn}=6.0 (\textit{IQR}=4.8-7.0)) conditions.
A Friedman test yielded a difference between the three conditions ($\chi^2(2) = 8.4$, \textit{p}=.01); post-hoc tests found a significant difference between Qlarify and Manual (\textit{p}=.02).
Participants also felt they were able to gain more information during their exploration in the Qlarify condition (5.8 (\textit{SD}=0.8), \textit{Mdn}=6.0, \textit{IQR}=(5.0-6.0)) than in Manual (3.8 (SD=1.4), \textit{Mdn}=4.0 (\textit{IQR}=3.5-5.0)) and QA (5.1 (\textit{SD}=1.4), \textit{Mdn}=5.0 (\textit{IQR}=4.0-6.0)); $\chi^2(2) = 9.6$, \textit{p}=.01; post-hoc tests found a significant difference between Qlarify and Manual (\textit{p}=.02).

Reinforcing observations of \sys{} scaffolding the question asking process, participants reported greater motivation to explore deeper into papers in the Qlarify condition (6.3 (\textit{SD}=1.1), \textit{Mdn}=7.0 (\textit{IQR}=5.8-7.0)) than in Manual (4.2 (\textit{SD}=0.6), \textit{Mdn}=4.0 (\textit{IQR}=4.0-4.3)) and QA (5.8 (\textit{SD}=1.4), \textit{Mdn}=6.0 (\textit{IQR}=4.8-7.0)); $\chi^2(2) = 15.4$, \textit{p}=.0004; post-hoc tests found a significant difference between Qlarify and Manual (\textit{p}=.002). Participants mentioned how \sys{} enabled them to retrieve information that felt ``detailed and useful,'' allowing them to understand for instance, ``what the authors had done, the authors' use of the terms they developed, the things that they built on'' (P8). In contrast, they found the QA condition sometimes delivered information that was ``too high-level,'' likely due to the more generic nature of questions they asked when unguided.

All participants appreciated the added value provided by \sys{} and reported a greater desire to use the Qlarify (6.3 (\textit{SD}=0.9), \textit{Mdn}=6.5 (\textit{IQR}=5.8-7.0)) condition compared to Manual (3.5 (\textit{SD}=1.5), \textit{Mdn}=4.0 (\textit{IQR}=2.8-4.0)) and QA (5.3 (\textit{SD}=1.7), \textit{Mdn}=6 (\textit{IQR}=4.5-6.3)); $\chi^2(2) = 10.2$, \textit{p}=.006; post-hoc tests found a significant difference between Qlarify and Manual (\textit{p}=.01).
Across all survey questions, participants reported measures that favored Qlarify over QA (Figure~\ref{fig:survey_likert_responses}), though none of these differences were significant after post-hoc correction for the multiple statistical tests in our study.

On average, we found participants viewed a similar number of abstracts across the three conditions (Manual: 6.0 (\textit{SD}=3.4), QA: 6.8 (\textit{SD}=3.5), Qlarify: 6.3 (\textit{SD}=2.5)), but opened fewer paper PDFs in the QA and Qlarify conditions (Manual: 5.3 (\textit{SD}=3.2), QA: 4.0 (\textit{SD}=1.7), Qlarify: 3.7 (\textit{SD}=1.8)). Furthermore, participants spent less total time in the paper PDFs in the Qlarify condition (Manual: 538.8s (\textit{SD}=223.5s), QA: 262.1s (\textit{SD}=97.2s), Qlarify: 201.1s (\textit{SD}=96.7s)). Together, these findings suggest participants were less inclined to view the full paper when provided with some affordances for information retrieval in the abstract, and spent less time skimming through the full PDF when an expandable abstracts were available. 

Lastly, in the Qlarify condition, we found 10 of 12 participants drilled down to view attributed evidence for an expansion highlighted in the PDF at least once during the study. When participants opened the PDF via attribution, they spent on average 32.4s (\textit{SD}=21.0s) in the PDF. In contrast, whenever participants opened the PDF from the title, they spent longer on average scrolling through the PDF, 55.6s (\textit{SD}=30.0s). These findings suggest participants used the attributed evidence as a quick entry point into the full paper, often as a ``guide back to the paper'' to ``cross-check whatever the AI generated.'' (P9)

\section{Evaluation of Generated Expansions}
We conducted a small-scale evaluation of the quality of expansions generated within \sys{}, using a subset of the data collected in the deployment study. Members of the research team annotated 120 randomly sampled expansions for which an answer was found (30 for each of the \textit{Define}, \textit{Expand}, \textit{Why}, and AI-suggested questions).

Of the 120 expansions, 105 (87.5\%) were entirely accurate (i.e., all statements were grounded in verifiable information from the paper) corroborating perceptions of high expansion quality by scholars in our interview study. While we did not explicitly assess relevance, our annotation and participant observations from our interview study suggest that the LLM-generated expansions were largely relevant to the queried information. Two of the main sources of error within the analyzed expansions included:

\textbf{Inaccurate details.} Seven (5.8\%) expansions included detail inaccuracies, often involving numerical or mathematical content. These included false navigational references (e.g., attributing statements to an incorrect section in the paper), numerical values in experimental results (e.g., fabricated numbers in ``the additional mean overhead time of 0.47s is only 12.8\% of the average episode duration of 3.67s''), and acronyms (e.g., describing FMs as ``language guided models'' rather than ``foundation models'').


\textbf{Missing content.} Eight (6.7\%) expansions contained phrases such as, ``the paper does not provide explicit details for\ldots .'' This error tended to occur when a portion of the relevant information was provided in a table rather than in the body text of the paper. Rather than a limitation of LLMs, this perhaps reflects how \sys{} preprocesses papers into a flat representation without delimitation for structural or visually salient content such as tables. In other cases, the context provided to the LLM may have lacked sufficient information to answer the question, suggesting the need to further investigate robust chunk retrieval techniques.

Overall, our analysis highlights the infrequent yet subtle errors within retrieval-augmented generation approaches, such as how plausible yet hallucinated details can become embedded into an otherwise accurate expansion. It is worth noting that no scholars in our interview study explicitly noticed any errors, and we identified errors only through extensive checking with the original paper. These hallucinations can thus be challenging to detect---especially within cognitive demanding or rapid exploratory processes such as triage---and potentially lead to harmful misinterpretations and an erosion of trust in the reliability of the generated expansions.

\section{Discussion} \label{sec:discussion}
In this paper, we describe \textit{recursively expandable abstracts}, a novel document-centric interaction technique that dynamically elaborates on abstracts along directed information needs with details retrieved from the full texts of papers. Findings from our studies reveal how \sys{} can support the retrieval of information from a paper on-demand through one-click questions, bridging the informational gaps between an abstract and a paper. In this section, we discuss potential extensions of \sys{} and opportunities for future generalizations of LLM-powered, interactive expansions.

Our exploration of a fluid text medium within \sys{} leaves exciting design dimensions of the interaction for future research. As mentioned in \S\ref{sec:designSpace}, abstracts could also be expanded for information needs not grounded in the abstract's text. For instance, expanding abstracts with agnostic information needs, i.e., questions that exist divorced of any specific abstract (e.g., ``What are their contributions''), could draw on prior work extracting key faceted information, e.g., contributions, method, and findings~\cite{fok_scim_2023, chan_solvent_2018}.

Some study participants also acknowledged the value in freely asking any question to a paper, similar to the QA condition of the comparative evaluation. While the AI-suggested expandable entities in \sys{} offer structured guidance toward potentially relevant information,  allowing users to freely ask their own questions could encourage a more self-driven and critical exploration of the paper. Indeed, we see the two interactions as complementary and addressing different types of information needs; future systems may provide users with both affordances for greater flexibility.


There are also opportunities to better support context switching between the reading of abstracts and full papers. For instance, systems could include additional entry points into a paper from the abstract. \sys{} provides one per expansion by allowing scholars to drill down into the paper's context though attributed evidence, which we saw helped participants validate and build trust in the AI-generated expansions. For an expansion, systems could instead help guide a scholar's attention to multiple related or complementary passages throughout the paper. An exciting direction is then exploring how these systems can help scholars retain their newly acquired knowledge as they return from the full paper back to the abstract triaging process. Participants in our studies used the ability to drill down into a paper not only to validate the generated expansions, but also as a means to quickly open the paper to browse visual content such as a system diagram or a table of experimental results. Rather than the text-only modality of \sys{}, future expandable abstracts may learn to surface graphical content from papers, such as figures and tables, which participants often desire in their exploration. Expandable abstracts may further benefit from personalization, with systems learning from a user's expansion history to suggest tailored expandable entities for new abstracts to lower interaction costs and encourage exploration, or even regenerate abstracts tailored to a user's interests.

Although \sys{} was designed and evaluated with papers largely within computer science, we believe the underlying interaction paradigm can transcend disciplines. For instance, prior work has shown that reading medical literature can be challenging and overwhelming due to barriers such as dense and unfamiliar terminology, not knowing what to read, and the inability to find answers to specific questions~\cite{august_paperplain_2023}. One avenue for future work could investigate how expandable abstracts with similar question asking and answering assistance could make medical papers more accessible by providing just-in-time responses to questions directly within an abstract, without requiring lay readers to interact with the full paper. A similar idea might allow policymakers to understand the implications of scientific developments, a critical need~\cite{Tyler2023AITA}. Future studies could also examine the potential for recursive expansion interactions over summaries of documents in other domains (e.g., legal documents, medical notes, or discussions in online forums). Moving beyond comprehension of a single paper and into synthesis across papers, another direction could explore support for expansion contexts with multiple documents. For example, how could \sys{} be extended to allow users to expand on related work sections and explore information across many cited papers, synthesizing the information retrieved across these multiple documents?

\subsection{Limitations}
While many participants appreciated the AI-suggested expandable entities, some felt its visual salience could hinder reading of the underlying abstract. Participants also found the in-situ expansions could interrupt their reading flow, especially if the generated answers were verbose or inaccurate. To mitigate these issues, future systems could provide customizability for the presentation (e.g., fluid text, sidebar, popup) and quantity (e.g., number of expandable entities) of AI-enabled reading augmentations. 

Beyond user control, the development of scholarly support tools also necessitates consideration of the implications of deploying such systems. While LLM-generated questions within \sys{} can help scaffold an active reading process, they also reduce a scholar's agency over their exploration. Interactions such as abstract expansions could disincentivize scholars from reading full papers, instead encouraging more superficial exploration through interactions with abstracts only. Several participants in our studies noted how over-reliance on the ``path of least resistance'' offered by such tools could undermine the knowledge and self-actualization gained from years of triaging and consuming research literature by themselves, and potentially harm new scholars in particular. Nonetheless, we believe these tools can significantly enhance the efficiency and efficacy of scholarly processes, though they should be developed mindful of these risks and pedogogical implications.

Finally, expandable abstracts (and many scholarly support tools) require access to a paper's full text or PDF.
In this work, we had the privilege of institutional access to the full text of many papers, but we note a significant portion of science remains inaccessible behind paywalls enacted by academic publishers. While legal and institutional challenges remain, open access initiatives (e.g., the Open Access movement\footnote{cf. \url{https://www.doaj.org/}, \url{https://www.budapestopenaccessinitiative.org/read/}}, arXiv, S2ORC~\cite{s2orc})) have made notable strides in changing the landscape of publishing to encourage more accessible dissemination of scientific knowledge.

\section{Conclusion}
This paper introduces \textit{recursively expandable abstracts}, a novel interaction technique that allows scholars to directly expand abstracts with clarifying information from paper full texts, and implements the interaction within an augmented reading interface, \sys{}. To support the creation of expansions, \sys{} leverages LLMs to identify entities within abstracts that are informative to expand, suggest intent-inferring questions that scholars can ask in a single click, and generate concise, attributed responses. Through an interview study, we found expandable abstracts helped scholars to rapidly and deeply address information-seeking needs during paper exploration. A field deployment further characterized how scholars would use expandable abstracts for paper triage in a real-world setting. Our final (comparative) evaluation showed that participants felt they could explore more deeply and were more satisfied with \sys{} than with plain abstracts, even if question answering was available. We believe \sys{} contributes a valuable step toward LLM-enabled systems that effectively empower an interactive, low-cost, and just-in-time exploration of long, complex texts.

\begin{acks}
\end{acks}

\bibliographystyle{ACM-Reference-Format}
\bibliography{zotero,aug}

\appendix
\section{Evaluation Details}

\subsection{Interview Questions} \label{appendix:interview-questions}
The following questions were used to elicit qualitative insights from participants in the interview study. These questions were used to initially guide the discussion, and probing questions were used to further elaborate on responses.
\begin{itemize}
    \item Summarize your overall impressions of this interactive abstract interaction. What did you like or dislike?
    \item What other features or improvements would you want in a future system with interactive abstracts?
    \item Did you prefer asking questions by highlighting text or by clicking one of the underlined entities, and why?
    \item How do you feel about the choice of three static questions provided in the question palette? Are there other questions you would have wanted?
    \item Did you use either the attributed evidence paragraph or the ability see the evidence in the context of the paper? If so, how did you use it and was it useful?
    \item How did you feel about the quality of the generated responses?
    \item How did you feel about the quality of the suggested question in the question palette?
    \item Do you think this interactive abstract would be useful in your own research workflows, and if so, how? 
\end{itemize}

\begin{figure*}[ht]
    \centering
    \includegraphics[width=0.95\textwidth]{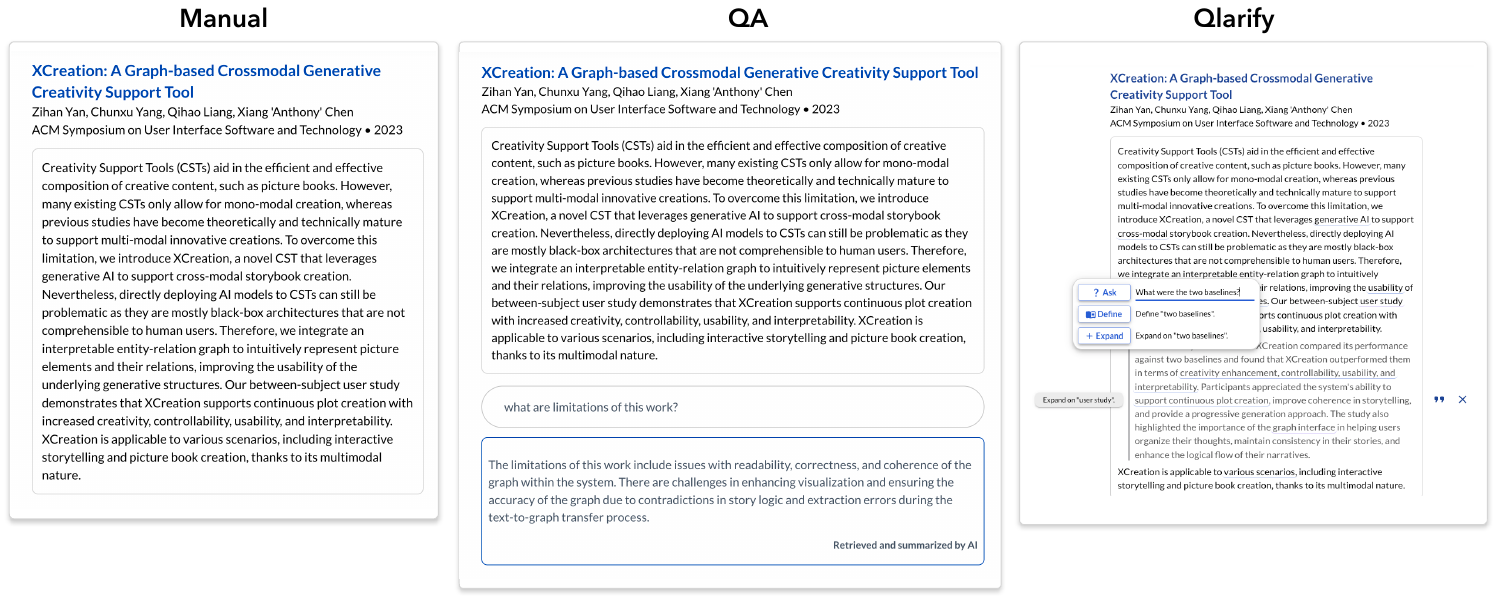}
    \caption{Comparison of systems in the three conditions in Study 3: Comparative Evaluation.}
    \Description[]{}
    \label{fig:study3_conditions}
\end{figure*}

\subsection{Post-Task Survey Questions} \label{sec:comp_eval_survey_questions}
For the post-task survey in the comparative evaluation, participants rated their agreement with the following statements on a seven-point Likert scale (1 = Strongly Disagree, 7 = Strongly Agree).
\begin{itemize}
    \item \textit{Satisfaction}: ``I felt satisfied with my experience using the tool to explore the abstracts/papers.''
    \item \textit{InfoGain}: ``The tool helped me gain relevant information while reading the abstracts/papers.''
    \item \textit{InfoAccurate}: ``The information gained through my exploration was accurate.''
    \item \textit{InfoReliable}: ``The information gained through my exploration was reliable.''
    \item \textit{Confidence}: ``I felt confident while exploring the abstracts/papers.''
    \item \textit{ExploreDeeper}: ``I felt motivated to ask questions or explore deeper into abstracts/papers.''
    \item \textit{ValueAdd}: ``I believe the tool can add value to my research process.''
    \item \textit{FutureUse}: ``If publicly available in the future, I would use a similar tool for exploring abstracts/papers.''
\end{itemize}

\subsection{System Updates for Comparative Evaluation}
\begin{figure*}[ht]
    \centering
    \includegraphics[width=0.95\textwidth]{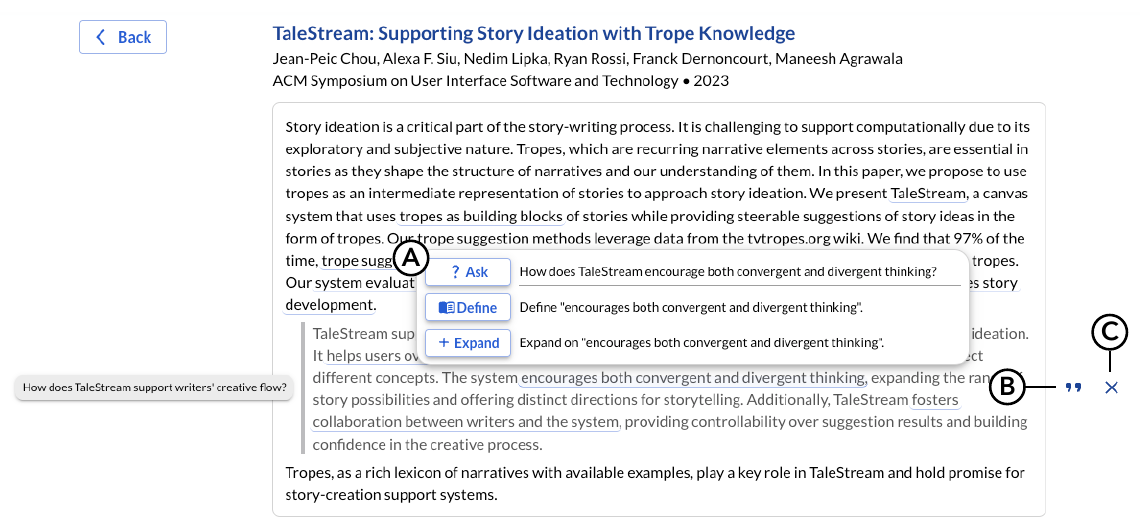}
    \caption{Updated user interface of \sys{} for the comparative evaluation. Refinements included allowing users to edit the AI-suggested question (A), moving the button for viewing attributed evidence in the PDF to the right margin (B), and adding a button to remove a generated expansion (C). }
    \Description[]{}
    \label{fig:study3_qlarify_ui}
\end{figure*}

Figure~\ref{fig:study3_qlarify_ui} illustrates the modifications made to \sys{}'s design based on feedback from the interview and field deployment studies.

\subsection{User Action Maps}
\begin{figure*}[ht]
    \centering
    \includegraphics[width=0.95\textwidth]{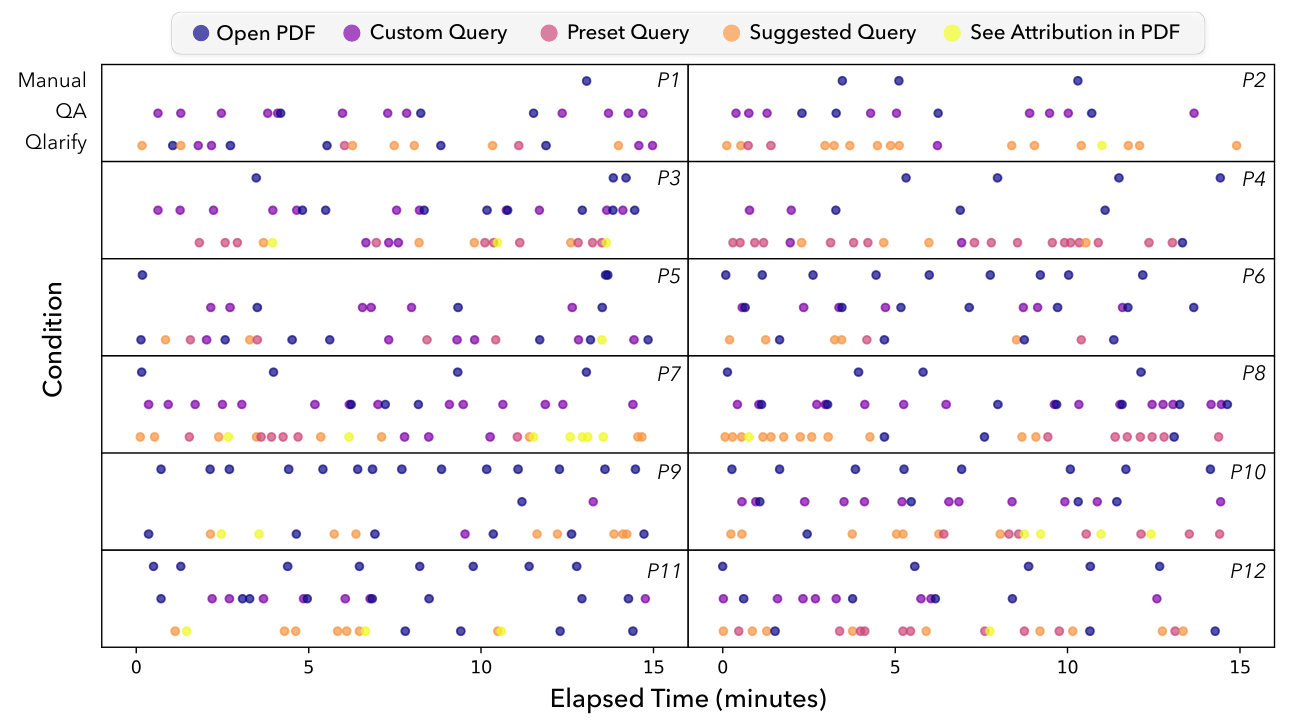}
    \caption{Time-graph of user actions during task completion in the comparative evaluation.}
    \Description[]{}
    \label{fig:user_action_maps}
\end{figure*}

Figure~\ref{fig:user_action_maps} illustrates how scholars used \sys{}'s features to explore paper abstracts, compared to baseline approaches of manual triage and standard question answering.

\section{Example \sys{} Outputs} \label{appendix:example-pipeline-outputs}
Table~\ref{tab:expansion-examples} lists LLM-generated questions and expansions for each detected expandable entity, and Table~\ref{tab:static-question-examples} lists example expansions for the static questions.

\begin{figure*}[ht]
    \centering
    \includegraphics[width=0.7\textwidth]{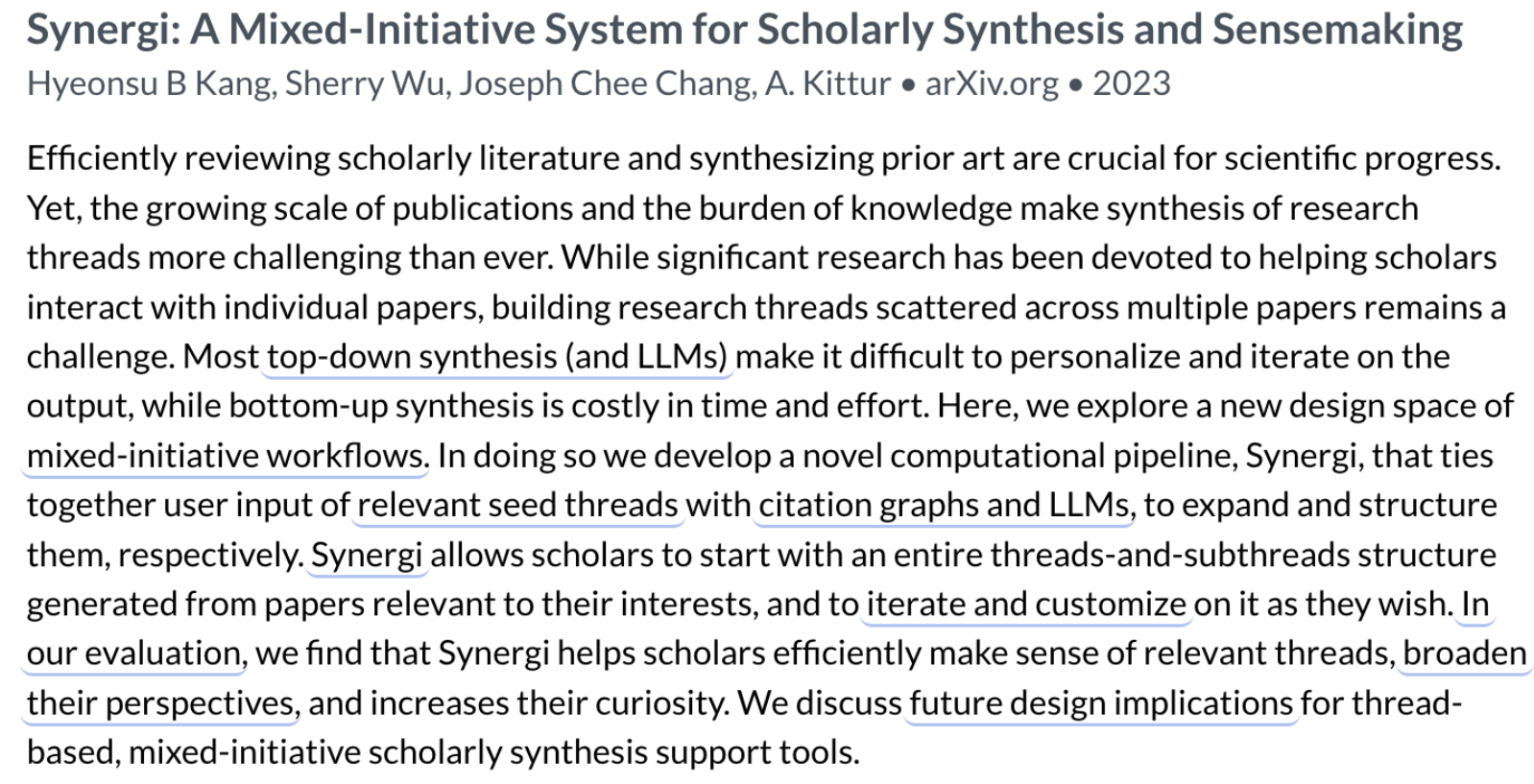}
    \caption{\sys{}'s user interface with expandable entities underlined and no expansions created (abstract shown is from~\cite{kang_synergi_2023}).}
    \Description[]{}
    \label{fig:synergi-abstract}
\end{figure*}
\begingroup

\renewcommand{\arraystretch}{1.2}

\begin{table*}[ht]
    \small
    \centering
    \caption{Examples of LLM-generated expansions for the three static questions in \sys{} (from~\cite{kang_synergi_2023}; abstract in Figure~\ref{fig:synergi-abstract}).}
    \begin{tabular}{p{0.1\textwidth}p{0.22\textwidth}p{0.55\textwidth}}
        \toprule
        \textbf{Question} & \textbf{Entity} & \textbf{Expansion} \\
        \midrule
        \textit{Define} & ``mixed-initiative workflows'' & Mixed-initiative workflows refer to a design approach that combines both human input and automated assistance in the process of conducting literature reviews and synthesizing knowledge from scientific papers. It involves the integration of user input, such as relevant seed threads, with computational algorithms and interaction features to expand and structure the threads. This approach allows scholars to start with a pre-generated structure and customize it according to their needs, facilitating efficient sensemaking, broadening perspectives, and increasing curiosity. \\
        
        \textit{Expand} & ``broaden their perspectives'' & Participants in the study engaged with Synergi-generated threads of research to "broaden their perspectives." This means that the threads provided them with a wider range of information and insights, allowing them to consider different viewpoints and explore various aspects of the literature. By broadening their perspectives, participants were able to gain a more comprehensive understanding of the topic they were reviewing. \\
        
        \textit{Why} & ``building research threads scattered across multiple papers remains a challenge'' & Building research threads scattered across multiple papers remains a challenge because it requires synthesizing information from multiple sources and organizing them in a coherent manner. Prior approaches have supported largely manual bottom-up synthesis workflows, which can be time-consuming and inefficient. Synergi addresses this challenge by automatically retrieving important papers from a 2-hop neighborhood on the citation graph and summarizing them in a hierarchical manner, allowing users to quickly discover relevant threads and understand them through synthesis by other scholars. \\
            
        \bottomrule
    \end{tabular}
    \label{tab:static-question-examples}
\end{table*}

\endgroup
\begingroup

\renewcommand{\arraystretch}{1.3}

\begin{table*}[ht]
    \small
    \centering
    \caption{Examples of \sys{}'s LLM-generated expandable entities, suggested questions, and expansions. All examples are actual system output for the paper \textit{Synergi: A Mixed-Initiative System for Scholarly Synthesis and Sensemaking}~\cite{kang_synergi_2023} (abstract in Figure~\ref{fig:synergi-abstract}).}
    
    \scalebox{0.95}{
    \begin{tabular}{p{0.22\textwidth}p{0.22\textwidth}p{0.49\textwidth}}
        \toprule
        \textbf{Entity} & \textbf{Question} & \textbf{Expansion} \\
        \midrule

        ``top-down synthesis (and LLMs)'' & What is meant by ``top-down synthesis'' and ``bottom-up synthesis''? & Top-down synthesis refers to the approach of using large language models (LLMs) to generate summaries or syntheses of research papers. It is characterized by limited personalization and iteration on the output. On the other hand, bottom-up synthesis involves practices such as citation chasing and footnote chasing to discover important papers related to a research problem. It is more time-consuming but allows for a more comprehensive synthesis. \\
        
         ``mixed-initiative workflows'' & What is the definition of a ``mixed-initiative workflow''? & A mixed-initiative workflow is a type of workflow that combines both human input and automated processes. It involves user input of relevant seed threads, which are then expanded and structured using citation graphs and LLMs. Scholars can start with a pre-generated structure and customize it as they wish, allowing for personalization and iteration. \\
         
         ``relevant seed threads'' & What is meant by ``relevant seed threads''? & ``Relevant seed threads'' refers to the structured research threads that are generated based on the selected clips and their associated citation contexts and seed references. These threads are considered relevant because they align with the user's interests and provide valuable information about specific research areas in the literature. \\
         
         ``citation graphs and LLMs'' & How does Synergi use citation graphs and LLMs to expand and structure research threads? & Synergi uses citation graphs to retrieve important papers from a 2-hop neighborhood and summarize them in a hierarchical manner. It generates a 3-level hierarchy of research threads by capturing the core commonalities among the papers. It also utilizes LLMs like GPT4 to condense scholarly text and provide synthesis. \\
         
         ``iterate and customize'' & What is the process for scholars to iterate and customize the threads-and-subthreads structure? & Scholars can review the system-generated threads, curate the ones that interest them, and build upon them iteratively. They can easily curate useful threads, references, and contexts into an editor using drag-and-drop. They can also prioritize references by using the group-by-reference view. They can expand and collapse the structured thread recommendations to access the relevant citation contexts and source materials. \\
         
         ``In our evaluation'' & How was the evaluation of Synergi conducted? & The evaluation of Synergi involved 12 participants and domain experts. They compared Synergi to a baseline system, Threddy, and GPT4. The evaluation measured the quality of the outlines, efficiency in constructing them, and participants' perception of Synergi-generated threads and experience. The evaluation used experts' judgment, survey responses, qualitative data from interviews, and participants' interaction logs. \\
         
         ``broaden their perspectives'' & How does Synergi help scholars broaden their perspectives? & Synergi helps scholars broaden their perspectives by providing them with system-generated threads of research that they can quickly scan to understand how different sub-group structures map to the broader literature. This understanding allows scholars to orient their attention towards specific areas that align with their interests. \\
         
         ``future design implications'' & What are the future design implications for thread-based, mixed-initiative scholarly synthesis support tools? & The future design implications for thread-based, mixed-initiative scholarly synthesis support tools include incorporating LLMs as components in computational pipelines, exploring the design space of interaction designs that benefit users in discovering, prioritizing, extracting, organizing, and synthesizing knowledge, and conducting additional ablation studies to understand the contributions of each component in the pipeline. \\
         \bottomrule
    \end{tabular}
    }
    \label{tab:expansion-examples}
\end{table*}

\endgroup

\section{LLM Prompts} \label{appendix:llm-prompts}
Table \ref{tab:prompts} lists the LLM prompts used for the NLP services in \sys{}.

\begin{table*}[t]
    \centering
    \caption{Prompts used in \sys. \textbf{\{ \}} refers to a placeholder.}
    \scalebox{1}{
    \begin{tabular}{ll}
        \toprule
        \textbf{Task} & \textbf{Prompt} \\
        \midrule
        
        \thead{Expandable Entity \\ Extraction} & \texttt{\thead[l]{
        You are a helpful research assistant that asks questions about abstracts of scientific papers.\\\\
        List all questions that a curious reader might have after reading this abstract. These questions \\ must not be answerable given the abstract, but may be answerable given the full paper. These \\ questions could help clarify vague terms, define jargon, request for more detail, or ask for \\ justification. Each question should be short and not contain multiple sub-questions. Provide a \\ phrase (three words or less) verbatim from the abstract that motivates each question. \\\\
        Title: \textbf{\{Title\}}\\
        Abstract: \textbf{\{Abstract\}}\\\\
        \textbf{\{Examples\}} \\\\
        Questions:
        }} \\
        
        \midrule
        
        \thead{Question \\ Generation} & \texttt{\thead[l]{
        You are a helpful research assistant that predicts what question a reader might have.\\\\
        A reader has highlighted a span of text in the abstract. What is the most likely question they \\ could ask about the span? The question must not be answerable given the abstract, but may be \\ answerable given the full paper. The question may help clarify vague terms, define jargon, request \\ for more detail, or ask for justification. The question should be short and not contain multiple \\ sub-questions. Try framing the question as: How? Why? What? Such as?\\\\
        Abstract: \textbf{\{Abstract\}}\\
        Target span: ``\textbf{\{Entity\}}'', in the sentence ``\textbf{\{Sentence\}}''\\
        Question:
        }}\\
        
        \midrule
        
        \thead{Question \\ Answering} & \texttt{\thead[l]{
        You are a helpful research assistant that answers questions about scientific papers.\\\\
        Answer the question based on the following excerpts from the full text of the paper. Incorporate \\ quotes verbatim from the excerpts when relevant. If the question cannot be answered from the \\ provided context, reply ``No answer.'' Your answer should be \textbf{\{Response Length\}}.\\\\
        \textbf{\{Examples\}}\\\\
        Context: \textbf{\{Context\}}\\
        Question: \textbf{\{Question\}}\\
        Answer:
        }} \\
        
        \bottomrule
    \end{tabular}
    }
    \label{tab:prompts}
\end{table*}

\end{document}